%% file: main.tex
\documentclass[pmlr]{jmlr}

\usepackage{longtable}

\usepackage{booktabs}

\usepackage{placeins} 
\usepackage{xcolor}

\theorembodyfont{\upshape}
\theoremheaderfont{\scshape}
\theorempostheader{:}
\theoremsep{\newline}

\jmlrvolume{273}
\jmlryear{2025}
\jmlrworkshop{iRAISE Workshop at AAAI 2025}

\title[Spotting Early Issues in Computer Science Proposals]{AI Mentors for Student Projects:\\ Spotting Early Issues in Computer Science Proposals}

 \author{\Name{Gati Aher} \Email{gaher@cs.cmu.edu}\\
  \Name{Robin Schmucker} \Email{rschmuck@cs.cmu.edu}\\
  \Name{Tom Mitchell} \Email{tom.mitchell@cs.cmu.edu}\\
  \Name{Zachary C. Lipton} \Email{zlipton@cs.cmu.edu}\\
  \addr Carnegie Mellon University, Pittsburgh, PA 15213, USA}

\begin{document}

\maketitle

\begin{abstract}
When executed well, project-based learning (PBL) engages students' intrinsic motivation, encourages students to learn far beyond a course's limited curriculum, and prepares students to think critically and maturely about the skills and tools at their disposal.
However, educators experience mixed results when using PBL in their classrooms: some students thrive with minimal guidance and others flounder. Early evaluation of project proposals could help educators determine which students need more support, yet evaluating project proposals and student aptitude is time-consuming and difficult to scale.
In this work, we design, implement, and conduct an initial user study ($n = 36$) for a software system that collects project proposals and aptitude information to support educators in determining whether a student is ready to engage with PBL.
We find that (1) users perceived the system as helpful for writing project proposals and identifying tools and technologies to learn more about, (2) educator ratings indicate that users with less technical experience in the project topic tend to write lower-quality project proposals, and (3) GPT-4o's ratings show agreement with educator ratings.
While the prospect of using LLMs to rate the quality of students' project proposals is promising, its long-term effectiveness strongly hinges on future efforts at characterizing indicators that reliably predict students' success and motivation to learn.

\end{abstract}

\input{text/introduction}
\input{text/related_work}
\input{text/system_design}
\input{text/system_eval}
\input{text/discussion}


\bibliography{bibliography}

\input{text/appendix}

\end{document}

%% file: text/introduction.tex
\section{Introduction}
\label{sec:introduction}


Project-based learning puts students in charge of their own education, giving them the freedom to decide what to learn, what to create, and what success means for them. It is an approach that intends to mimic real life, where answers are not easy to find, and success requires curiosity, critical thinking, and resilience. PBL is frequently used in computer science classes because it encourages students to test their understanding against real-world problems, develop essential project management skills, and explore industry tools and practices outside of the standard classroom curriculum.


However, PBL has shown mixed effectiveness in the classroom, with learning and motivation depending on students' prior ability, educators' ability to support specific projects and topics \citep{Barnes2008:Increasing}, and whether the project idea originated from students or teachers \citep{pucher2011project}. Evaluating project proposals may help educators identify which students require additional guidance or alert educators to major issues that could consume limited motivation, time, and resources. Yet evaluating project proposals is challenging and time-consuming, requiring educators to review freeform text on a broad range of topics and infer student aptitude from incomplete or vague details.

In this work, we design, implement, and conduct a user study for a software system that collects project proposals and aptitude information that may help educators determine whether a student is ready to engage with PBL. Our design elicits proposals for a high school-level computer science project where students build and design an interactive web application or video game. Our project proposal form asks users to (1) describe a problem they want to work on, (2) propose a solution, (3) recall and analyze design inspirations, (4) predict the effects of their design, (5) plan to evaluate and iterate on their project, (6) describe skills they want to develop, and (7) connect those skills to computer science careers, technical tasks, and popular industry technologies.

Our preliminary user study ($n=36$) showed our system could engage users' intrinsic motivation, with $88.8\%$ of users wanting to use our system in the future to choose skills and technologies to learn more about, and $91.6\%$ of users wanting to use our system in the future to design project ideas that motivate them to learn more. Two educators and GPT-4o independently graded the quality of each project proposal according to a $29$-item rubric. Educators and GPT-4o showed promising agreement on the relative quality of proposals and tended to give lower quality scores to users with less computer science experience.

Our findings suggest that LLMs show promise for scaling the automatic grading of project proposals; however, the effectiveness of using LLM grades to guide instructional design decisions hinges on whether project proposals and grading criteria contain information that reliably predicts whether a student can benefit from project-based learning.

%% file: text/related_work.tex
\section{Related Work}
\label{sec:related_work}



\cite{Barnes2008:Increasing} found that student engagement in an Algebra 1 class improved when activities incorporated opportunities for choice, goal-setting, and one-on-one interactions with teachers. Similarly, \cite{pucher2011project} reported increased motivation among students when employing PBL in computer science classes. However, both studies highlighted challenges tied to students’ limited meta-cognitive skills and domain knowledge. For instance, many students struggled to maintain their goal portfolios \citep{Barnes2008:Increasing}, and student-defined projects received significantly lower grades from faculty \citep{pucher2011project}. \cite{kirschner2006unguided} critique PBL's effectiveness in supporting student learning, arguing that under minimal guidance, inexperienced students often struggle to develop effective strategies for independently searching for and applying information. Recent advances in LLMs enabled new types of learning technologies in various educational domains~\citep{Kasneci2023:Chatgpt}. LLMs' ability to provide feedback to student submissions~\citep{Botelho2023:Leveraging} and to support student self-reflection~\citep{Yazici2024:Gelex} inspires the present development of a system to support effective execution of PBL.

%% file: text/system_design.tex
\section{System Design}
\label{sec:design}



Our software system is implemented as an React.js and firebase-based web application that collects project proposals and aptitude information that may help educators determine whether a student is ready to engage with PBL. The system logs the time and type of user actions, such as keystroke patterns, answers to multiple-choice selections, and when users navigate away from and back to the web page. Our system first asks users questions about their aptitude for problem-solving, experience with computer science technologies, and prior experience with PBL. Then the system asks users to write down their idea for a high school-level computer science project on developing an interactive web application or video game. Our project proposal form, adapted from \cite{CSPathway, Byrdseed}, asks users to (1) describe a problem they want to work on, (2) describe the software they want to build, (3) recall and analyze design inspirations, (4) predict the effects of their design, (5) make a plan to evaluate and iterate on their project, (6) self-evaluate using a 10-point quality checklist, (7) describe skills they want to develop, and (8) connect those skills to computer science careers, technical tasks, and popular industry technologies. Due to space limitations, we included UI images in Appendix~\ref{app:user_interface}.

%% file: text/system_eval.tex
\section{System Evaluation}
\label{sec:evaluation}

Each project proposal was independently evaluated by two human domain experts with experience as head teaching assistant (TA) for undergraduate computer science classes and GPT-4o \citep{hurst2024gpt}. All raters used the same 23-item quality checklist with four subtasks: (1) 10-question quality checklist students had used to self-assess their own work, (2) 3 questions judging the quality of 3 skill descriptions written by students, (3) 9 questions judging the appropriateness of skill-career pairings ($3 \times 3$), and (4) an overall quality judgment question (``I would recommend a student include this project on their resume.”). We did not perform any prompt-engineering on GPT-4o beyond evaluating subtasks and each of the student's three written skill descriptions in a separate prompt to avoid holistic evaluations. We provide the evaluation rubric in Appendix \ref{app:expert_evaluation_rubric} and GPT-4o prompts in Appendix \ref{app:llm-prompts}.

This evaluation protocol captures two main signals of user readiness for PBL. First, by comparing users' self-assessments to expert human evaluations on the same 10-point \textbf{quality checklist} adapted from \cite{WPIRubric} and \cite{lawlor2012smart}, we could determine whether users can accurately report on the quality of their own work. 

Second, by asking human expert raters to classify the \textbf{quality of skill descriptions} (``Good" vs. ``Irrelevant", ``Vague", or ``Not Core Computer Science Skill") of three skills users want to develop by working on their project and the \textbf{appropriateness of how users matched those skills to predefined mentor profiles} (such as ``Data Scientist" or ``Software Developer"), career-specific tasks, and trending technologies sourced from the O*Net Online Database \citep{Onet2024:Online}, we could capture signal on whether users' are capable of identifying perspectives and types of industry knowledge relevant to their project ideas. Users were explicitly instructed to leverage internet resources for assistance if needed, so this metric was intended to measure general reasoning and searching ability rather than ability to recall specific definitions. However, misinterpreting instructions, inability to parse technical language used by O*Net Online Database \citep{Onet2024:Online}, and shoehorning skills to match one of a limited number of options could weaken the strength of this measurement. We sanity check the quality of our grading criteria by comparing scores from users with higher levels of computer science experience against users with less experience. We expect that students with computer science experience should be able to write higher quality project proposals.

Additionally, we asked the human experts raters and GPT-4o to answer a question meant to capture the overall quality of the project proposal (``I would recommend a student include this project on their resume. Yes/No").

We checked the motivational benefits of our project proposal writing activity with a post-activity experience survey.




\textbf{Experimental Procedure.} We recruited 40 participants online via Prolific. Our recruitment call was shown to crowd workers (i) 18 years or older; (ii) located in the USA; (iii) fluent in English; (iv) possess at least a high-school (HS) degree or equivalent; (v) using a desktop device (no mobile or tablets); and (vi) answered yes to ``Are you a student?" on Prolific (see recruitment call in Appendix \ref{app:prolific_recruitment}). 2 participants were filtered out for not completing the activity, and 2 were filtered for failing attention checks. Of the 36 remaining participants, 17 were 18-25 years old, 14 were 26-35 years old, and 5 were over 46 years old. 25 were male, 11 were female. 11 participants reported Computer Science as their field of work/study, and 19 claimed to have computer science experience, i.e, have built a website or interactive application before using ``Programming Language (e.g., Python, JavaScript)", ``Backend Technologies (e.g., Node.js, Django, Flask)", ``Cloud Platforms (e.g., Firebase, AWS)", and/or ``Testing and Deployment (e.g., unit testing libraries, Docker)". Interestingly, some of these users claimed experience in categories like Backend Technologies or Cloud Platforms while simultaneously reporting no experience with Programming Languages. This discrepancy suggests a potential need for more fine-grained and specific questions to accurately capture their technical experience. 

All participants encountered the same set of activities and questions in the same order, and could not go back to previous stages. Participants were told the activity would take 30-45 minutes, but we did not control for time in any phase. The median time to complete the activity was 40 minutes. The recruitment and study process was approved by CMU's Institutional Review Board (research study 2024\_00000405).

\subsection{Results}
As shown in Table \ref{tab:agreement}, human expert raters and GPT-4o showed promising agreement on the relative quality of proposals. As shown in Table~\ref{tab:positive_grades}, human expert raters and GPT-4o tended to give lower quality scores to users with less computer science experience.

\input{tables/agreement}

\input{tables/grade_table}

\textbf{Skill Classification.} Human raters and GPT-4o classifed ~50\% of skills as irrelevant, vague, or not relevant to computer science ($>80\%$, $\kappa>0.6$). Consistent across all three graders, students with less computer science knowledge had an average of 1 out of 3 skills accepted, while students with more computer science knowledge had an average of 2 of 3 skills accepted.

\begin{figure}[t]
\floatconts
  {fig:mean-score-spearman-corr}
  {\caption{Spearman correlation of students' total score on the 10-point quality checklist. Although GPT-4o's scores for the quality checklist task are lower than both teaching assistants and student self-evaluations, GPT-4o's scores preserve the rank order of teaching assistants' scores better than students' self-evaluation scores do.}}
  {\includegraphics[width=0.5\textwidth]{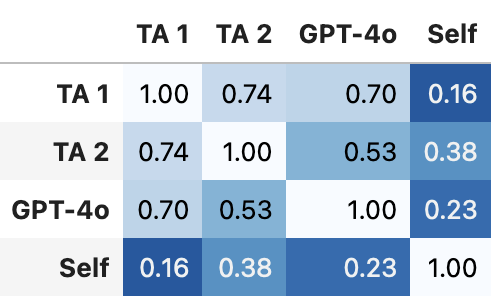}}
\end{figure}

\textbf{Appropriateness of Skill Pairing Classification.} The low quality of skills written down by the students might have made the subsequent activity to pair skills with careers, tasks, and technologies more challenging to grade. Human rater agreement was minimal ($68\%, \kappa=0.25$). GPT-4o showed weak agreement with TA 1 ($74\%, \kappa=0.46$) and minimal agreement with TA 2 ($67\%, \kappa=0.20$).

\textbf{Quality Checklist.}  GPT-4o’s scores tended to be lower than either human rater's scores, leading to lower agreement between GPT-4o and humans ($71-75\%, \kappa=0.28$). However, as seen in Figure \ref{fig:mean-score-spearman-corr}, GPT-4o roughly maintained the rank of student project proposal quality relative to each other, shown by GPT-4o’s Spearman correlations with TA 1’s (Spearman=0.70) and TA 2’s (Spearman=0.53). In contrast, students’ self-evaluations had a much weaker correlation with either human expert's ratings (Spearman=0.16, Spearman=0.38).

\textbf{Experience Survey.} As shown in Figure \ref{fig:experience}, the large majority of users enjoyed the process of writing project proposals, with $88.8\%$ of users wanting to use our system in the future to choose skills and technologies to learn more about, and $91.6\%$ of users wanting to use our system in the future to design project ideas that motivate them to learn more.

\begin{figure}[t]
\floatconts
  {fig:experience}
  {\caption{\textbf{Experience Survey, Response Counts.} The majority of users reported high level of excitement, motivation, and wanting to use the activity in the future to choose skills and technologies to learn more about.}}
  {\includegraphics[width=0.9\textwidth]{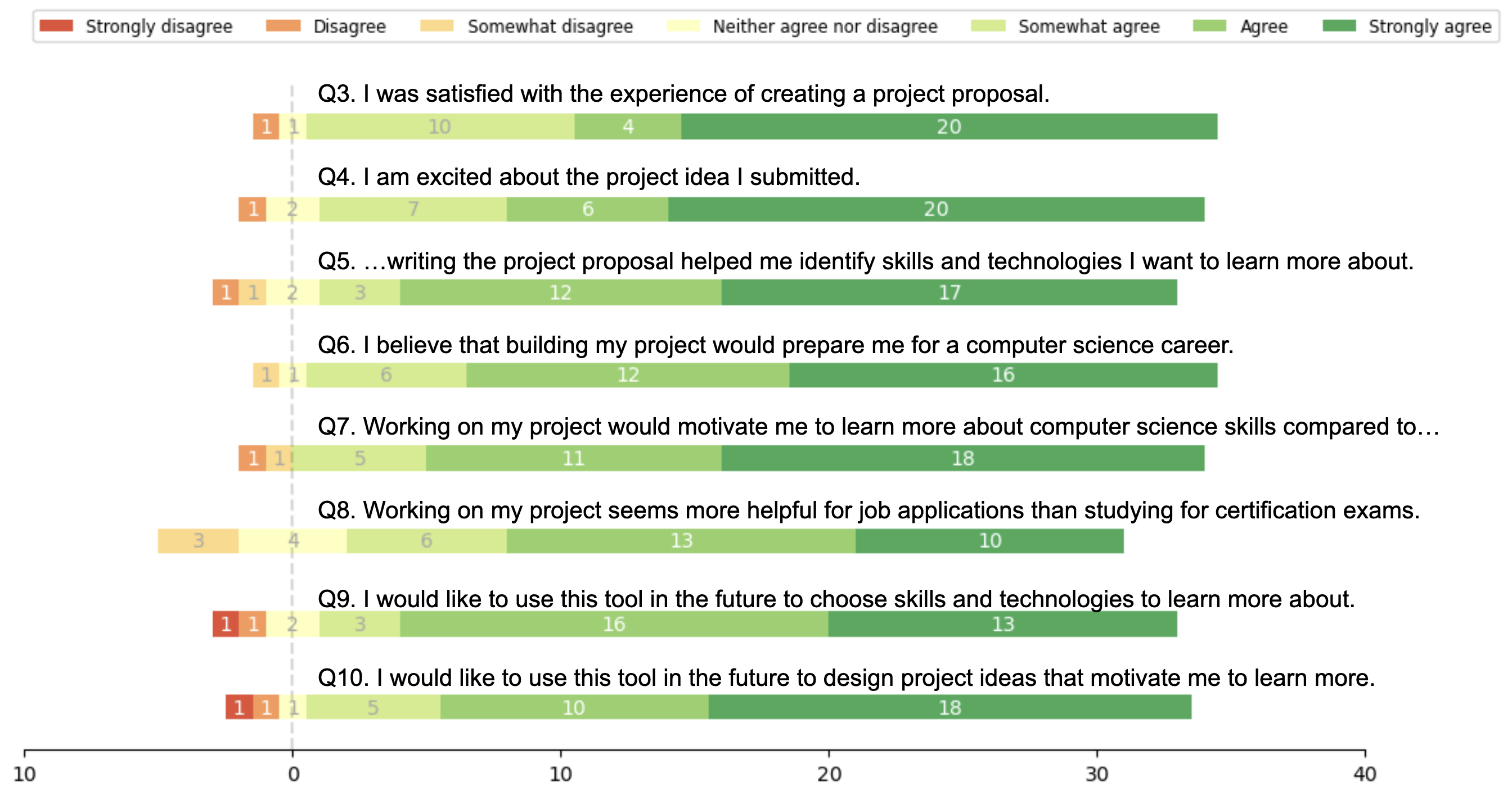}}
\end{figure}

%% file: tables/agreement.tex
\begin{table}[t]
\centering
\caption{Rater Agreement. We report agreement percentage and Cohen’s $\kappa$ value between the two human expert raters (TA1 and TA2) and GPT-4o with respect to the 4 grading subtasks. As an overall index of agreement, we compute kappa for all rater pairs then report the arithmetic mean of these estimates.
}
\begin{tabular}{lcccc}
\hline 
& TA1 / TA2 & TA1 / GPT-4o & TA2 / GPT-4o  & Avg. $\kappa$ \\
\hline
 Skill Quality Classification & $86.1\%$, $\kappa = 0.72$ & $84.3\%$, $\kappa = 0.68$ & $81.5\%$, $\kappa = 0.63$ & $0.68$ \\
 Skill Pairing Classification & $68.6\%$, $\kappa = 0.26$ & $74.2\%$, $\kappa = 0.46$ & $67.2\%$, $\kappa = 0.20$  & $0.29$ \\
 Quality Checklist& $87.2\%$, $\kappa = 0.49$ & $71.4\%$, $\kappa = 0.28$ & $74.7\%$, $\kappa = 0.28$ & $0.38$ \\
 Recommend for Resume & $80.6\%$, $\kappa = 0.50$ & $75.0\%$, $\kappa = 0.43$ & $77.8\%$, $\kappa = 0.45$ & $0.46$ \\
\hline
\end{tabular}
\label{tab:agreement}
\end{table}

%% file: tables/grade_table.tex

\begin{table}[t]
    \centering
    \caption{Mean Positive Grades for Experienced vs. Novice Students. Each cell represents the mean proportion of student responses graded positively per subtask and rater, grouped by whether students reported having previous computer science experience. Subtasks averaged over 3, 9, 10, and 1 rater items respectively}
    \begin{tabular}{rrcccc}
    \hline
     Task & Experience & TA1 & TA2 & GPT-4o & Self-Rating \\
    \hline
     Skill Classification & Novice & 35.3\% & 39.2\% & 31.4\% & - \\
     & Experienced & 70.2\% & 68.4\% & 68.4\% & - \\
     \hline
     Skill Pairing Classification & Novice & 60.8\% & 86.9\% & 57.5\% & - \\
     & Experienced & 53.7\% & 92.4\% & 64.9\% & -  \\
     \hline
     Quality Checklist & Novice & 83.5\% & 82.4\% & 58.2\% & 85.9\% \\
     & Experienced & 84.7\% & 90.0\% & 71.1\% & 95.8\% \\
     \hline
     Recommend for Resume & Novice & 58.8\% & 70.6\% & 58.8\% & - \\
     & Experienced & 78.9\% & 84.2\% & 73.7\% & - \\
    \hline
    \end{tabular}
    \label{tab:positive_grades}
\end{table}

%% file: text/discussion.tex
\section{Discussion and Limitations }
\label{sec:discussion}

Our findings suggest that LLMs show promise in scaling the automatic grading of project proposals; however, the effectiveness of using LLM grades to guide instructional design decisions hinges on whether project proposals and grading criteria contain information that reliably predicts whether a student can benefit from project-based learning. There are several limitations of our work that we now discuss.

First, in an optional feedback text field, human expert raters noted that several project proposals contained vague implementation details and skills, making it challenging for raters to determine whether crowd workers had appropriately paired their skills to computer science careers, tasks, or technologies (inter-TA agreement of 68.61\%, $\kappa=0.26$). To address these concerns, we gave raters a skill classification rubric to quantify the vagueness and irrelevance of skills, achieving moderate agreement among human expert raters (inter-TA agreement of 86.11\%, $\kappa=0.72$). Consistent across both human expert ratings and GPT-4o ratings, students with less computer science knowledge wrote an average of 1 out of 3 acceptable skills, while students with more computer science knowledge wrote an average of 2 of 3 acceptable skills. The effectiveness of project-based learning relies on students' recognizing the necessary skills to develop and maintaining motivation to learn them. While our current work focuses on leveraging LLMs to evaluate issues in project proposals, future iterations of our system may be able to improve students' success in project-based learning by helping beginners gain an awareness of specific and contextually relevant computer science skills.

Second, our study examines project proposals written by crowd workers who were 18 years or older and identified as students (answered yes to ``Are you a student?"). In this preliminary work, our aim was to gather project proposals from students with diverse backgrounds and varying degrees of familiarity with computer science. This approach attempted to capture the potential diversity of students that might enroll in introductory computer science courses at the high school or undergraduate level. The trends observed in the collected proposals can inform the design of future tools to support students in writing project proposals. Future work will investigate whether trends and issues observed in crowd workers' project proposals transfer to classroom assignments. 

Third, while generative AI-based learning technologies like our GPT-4o-based system show promise, they also incur regular costs due to API calls, raising important questions about equity and accessibility in educational contexts. While we use GPT-4o, we should be able to use our LLM-powered grading methodology with any generative large language model. Future work could explore the application of different LLMs for this task, including open-source models.

\section{Acknowledgements}

This work is supported by the Learning Engineering Tools Competition from The Learning Agency. We thank LeRoy Wong, Shereen Tyrrell, Minerva Sharma, Liz Chu, and the reviewers for insightful feedback.

%% file: text/appendix.tex
\appendix

\section{Illustrations of User Interface}
\label{app:user_interface}

Here, we provide additional details regarding the system workflow and user interface. The system guides users through a series of steps prompting the user to (1) describe a problem they want to work on (Figure~\ref{fig:s1} and~\ref{fig:s2}), (2) propose a solution (Figure~\ref{fig:s3}), (3) recall and analyze design inspirations (Figure~\ref{fig:s2}), (4) predict the effects of their design (Figure~\ref{fig:s4}), (5) plan to evaluate and iterate on their project, (6) describe skills they want to develop, and (7) connect those skills to computer science careers, technical tasks, and popular industry technologies (Figure~\ref{fig:s5},~\ref{fig:s6},~\ref{fig:s7},~\ref{fig:s8} and ~\ref{fig:s9}).

\begin{figure}[t]
\floatconts
  {fig:s1}
  {\caption{Introduction of project-based learning (PBL) task and possible projects.}}
  {\includegraphics[width=0.85\textwidth]{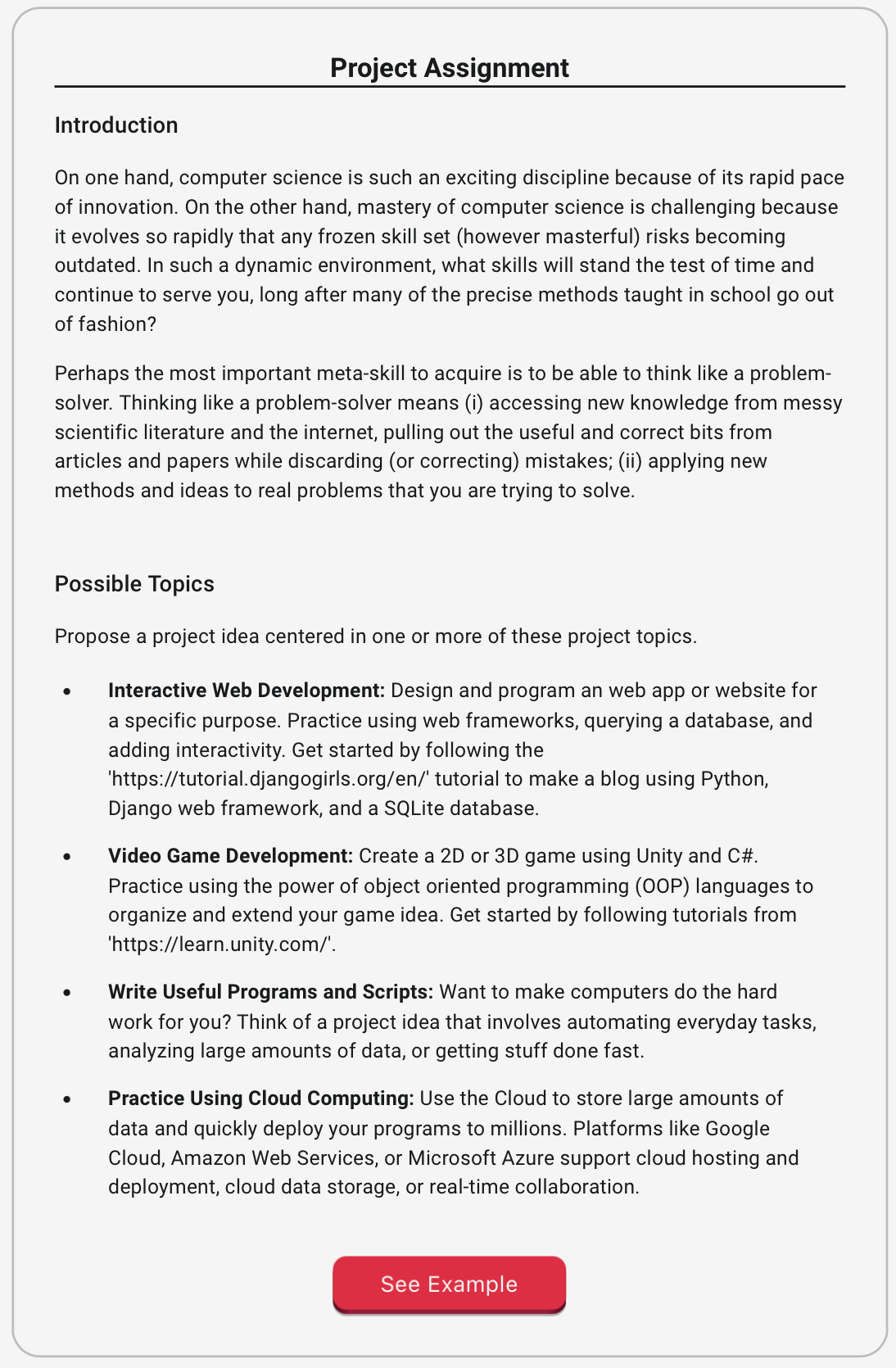}}
\end{figure}

\begin{figure}[t]
\floatconts
  {fig:s2}
  {\caption{In the first step, the system prompts the student to specify a project and describe background and objectives.}}
  {\includegraphics[width=0.82\textwidth]{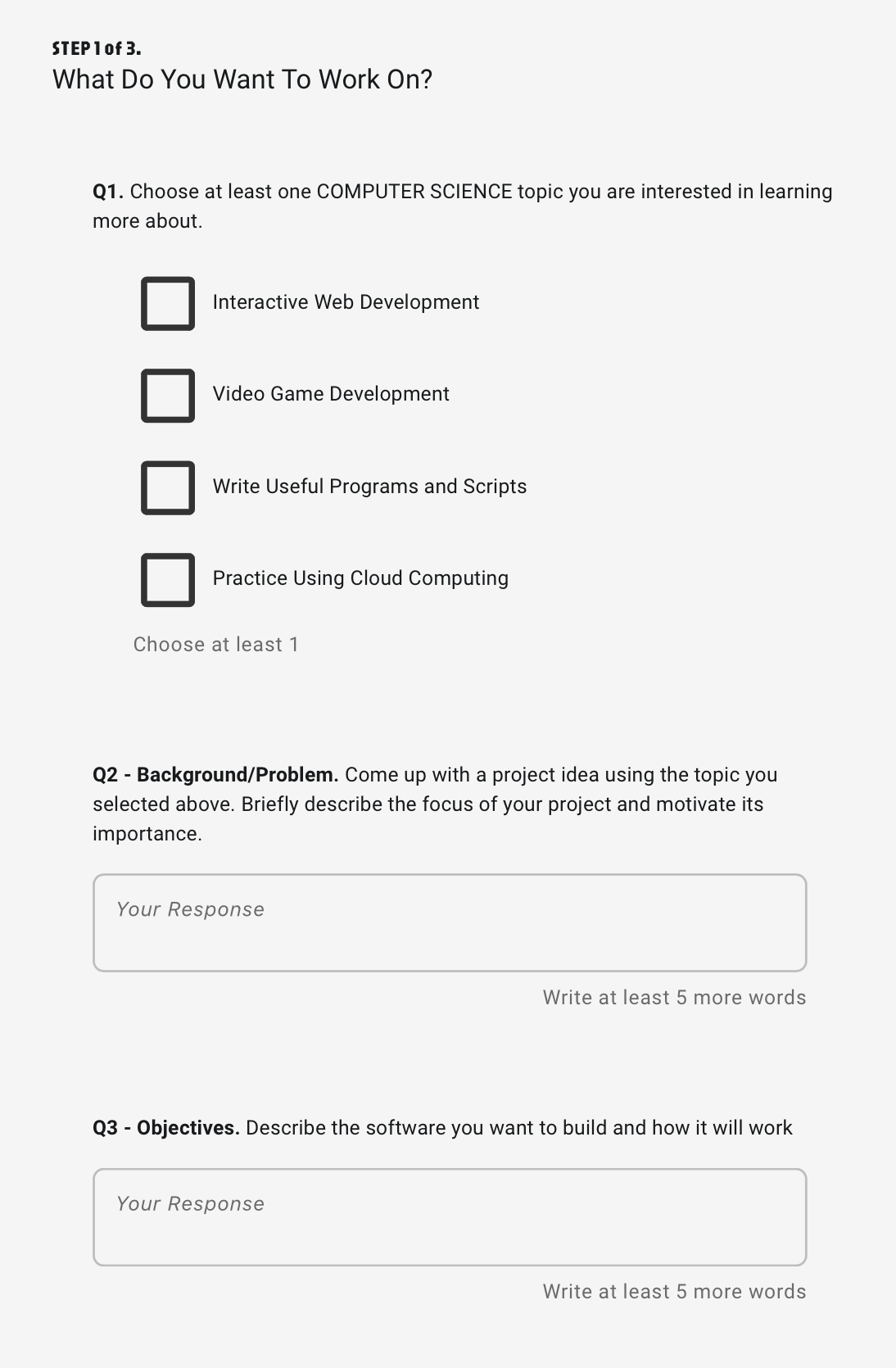}}
\end{figure}

\begin{figure}[t]
\floatconts
  {fig:s3}
  {\caption{In the second step, the system prompts the student to find inspiration and to connect the problem to existing examples.}}
  {\includegraphics[height=0.82\textheight]{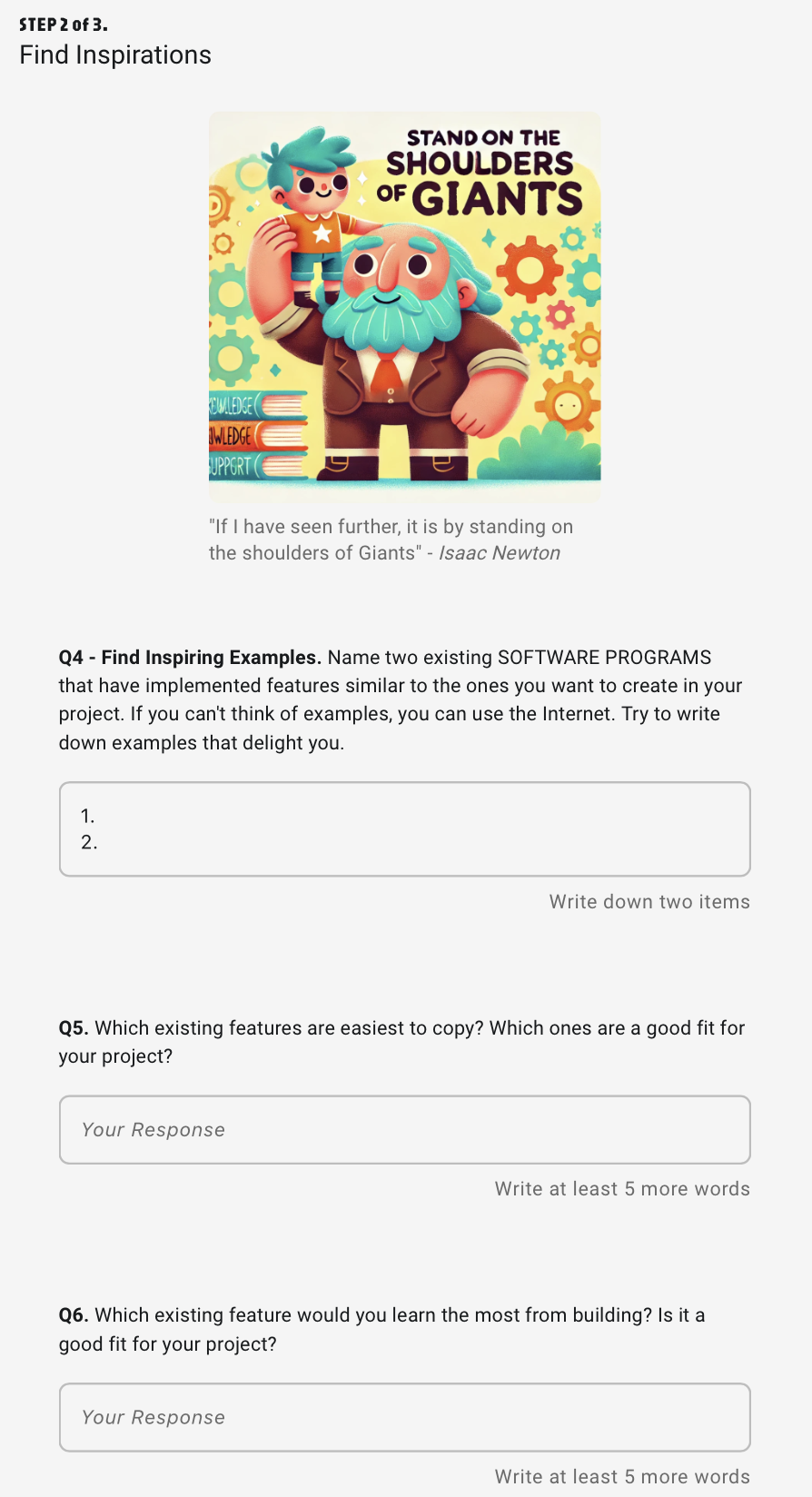}}
\end{figure}

\begin{figure}[t]
\floatconts
  {fig:s4}
  {\caption{In the third step, the system prompts the student to specify their design in greater detail.}}
  {\includegraphics[width=0.72\textwidth]{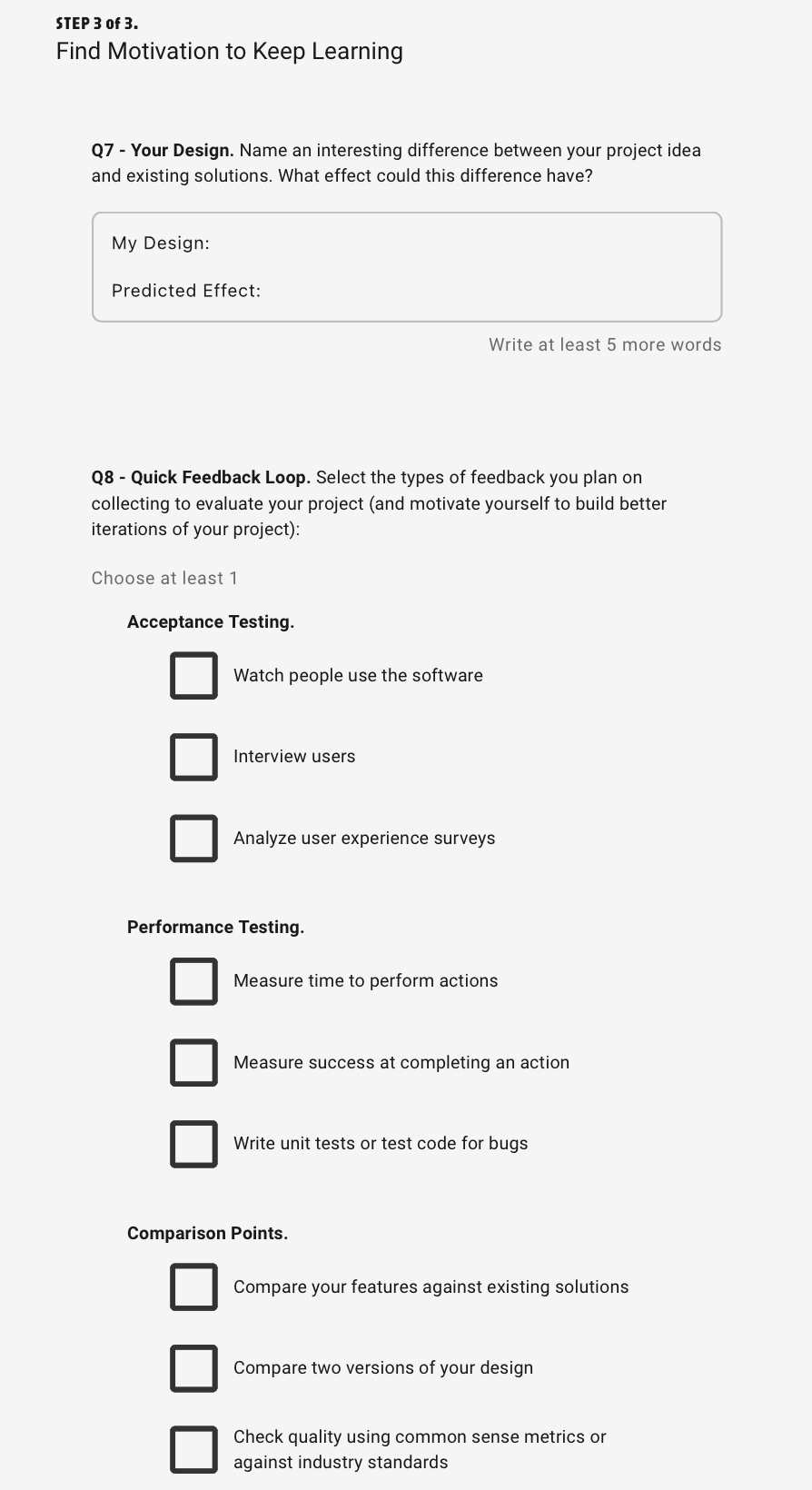}}
\end{figure}

\begin{figure}[t]
\floatconts
  {fig:s5}
  {\caption{The system prompts the student to reflect on their proposal in terms of background and problem.}}
  {\includegraphics[width=0.9\textwidth]{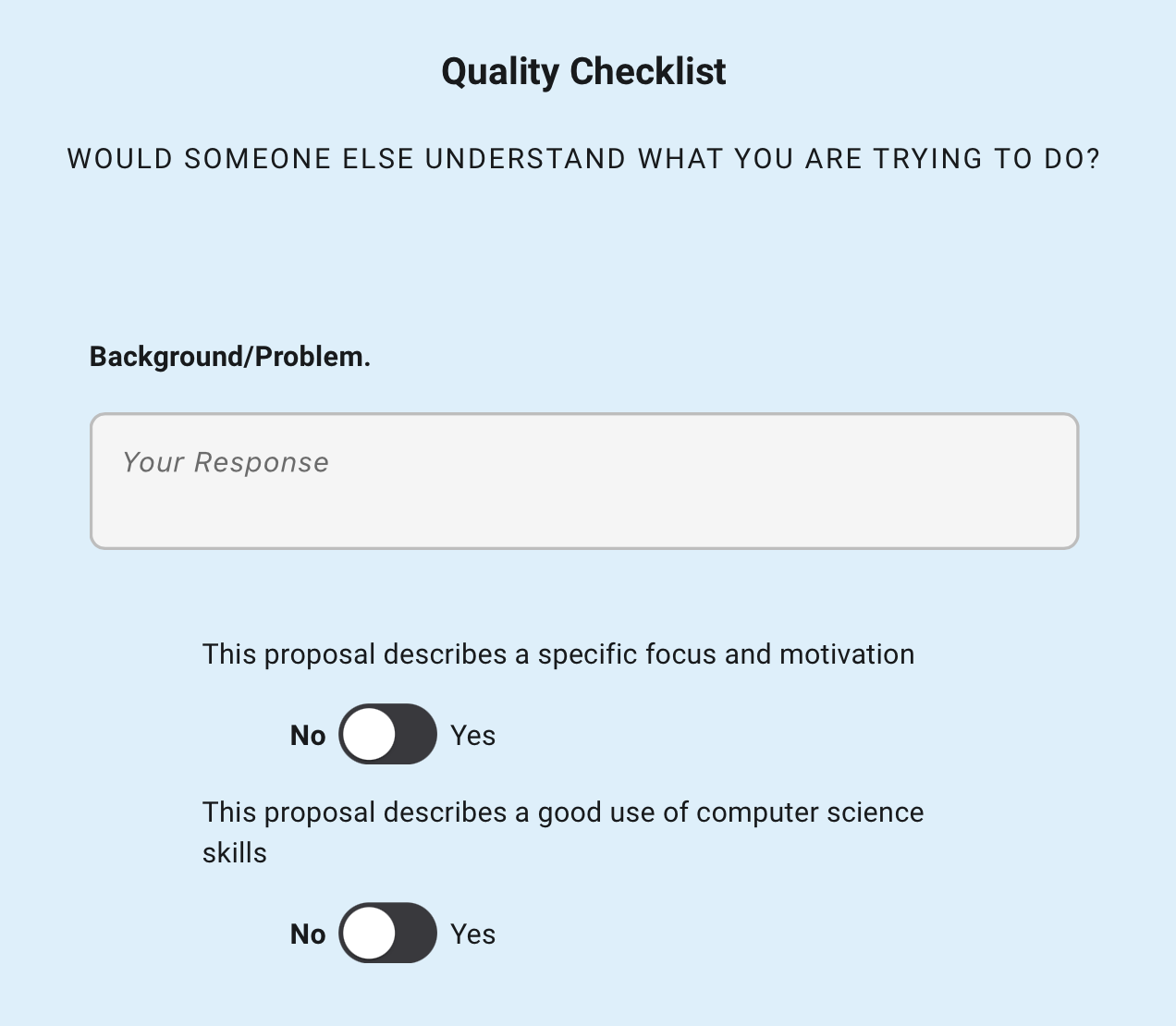}}
\end{figure}

\begin{figure}[t]
\floatconts
  {fig:s6}
  {\caption{The system prompts the student to reflect on their proposal in terms of project objectives. After answering Q1 (choosing topic), the second question in the checklist is populated with the chosen topic's name.}}
  {\includegraphics[width=0.9\textwidth]{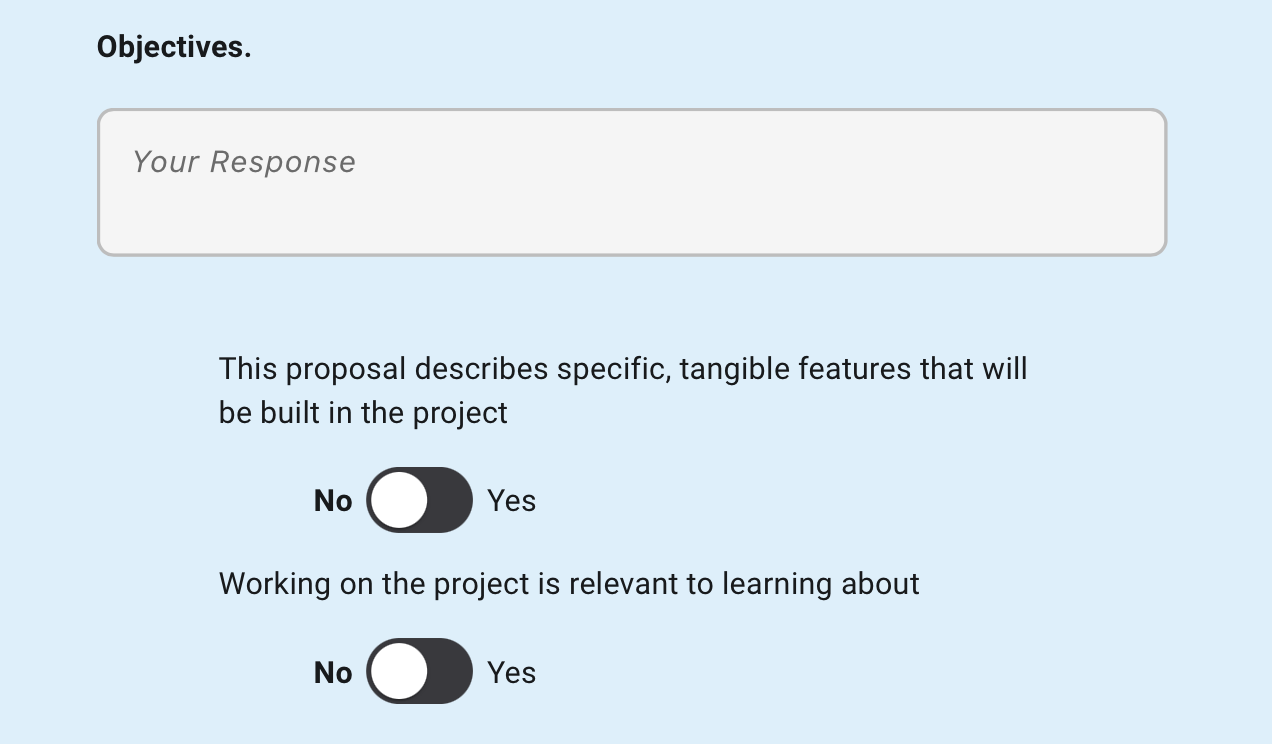}}
\end{figure}

\begin{figure}[t]
\floatconts
  {fig:s7}
  {\caption{The system prompts the student to reflect on their proposal in terms of related work and references.}}
  {\includegraphics[width=0.9\textwidth]{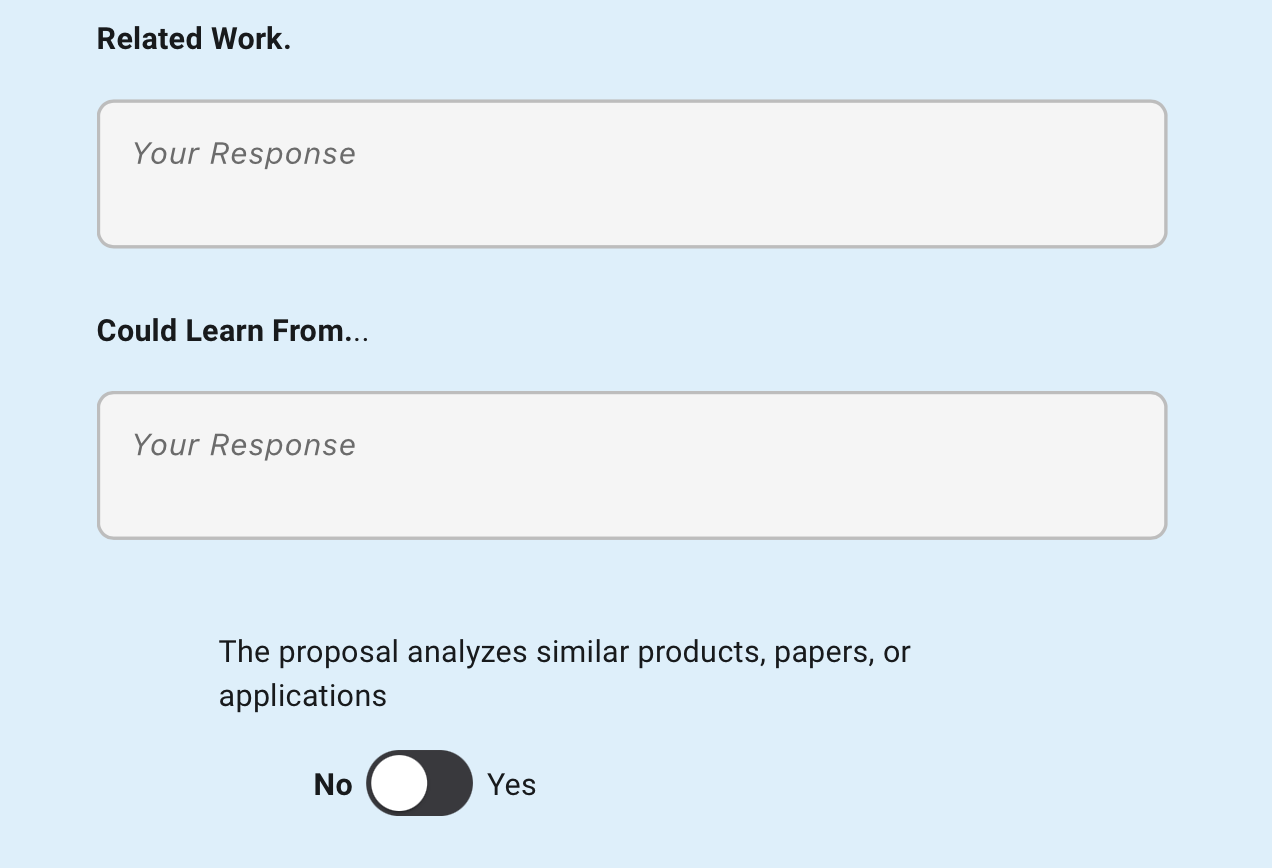}}
\end{figure}

\begin{figure}[t]
\floatconts
  {fig:s8}
  {\caption{The system prompts the student to reflect on their proposal in terms of design hypothesis.}}
  {\includegraphics[width=0.9\textwidth]{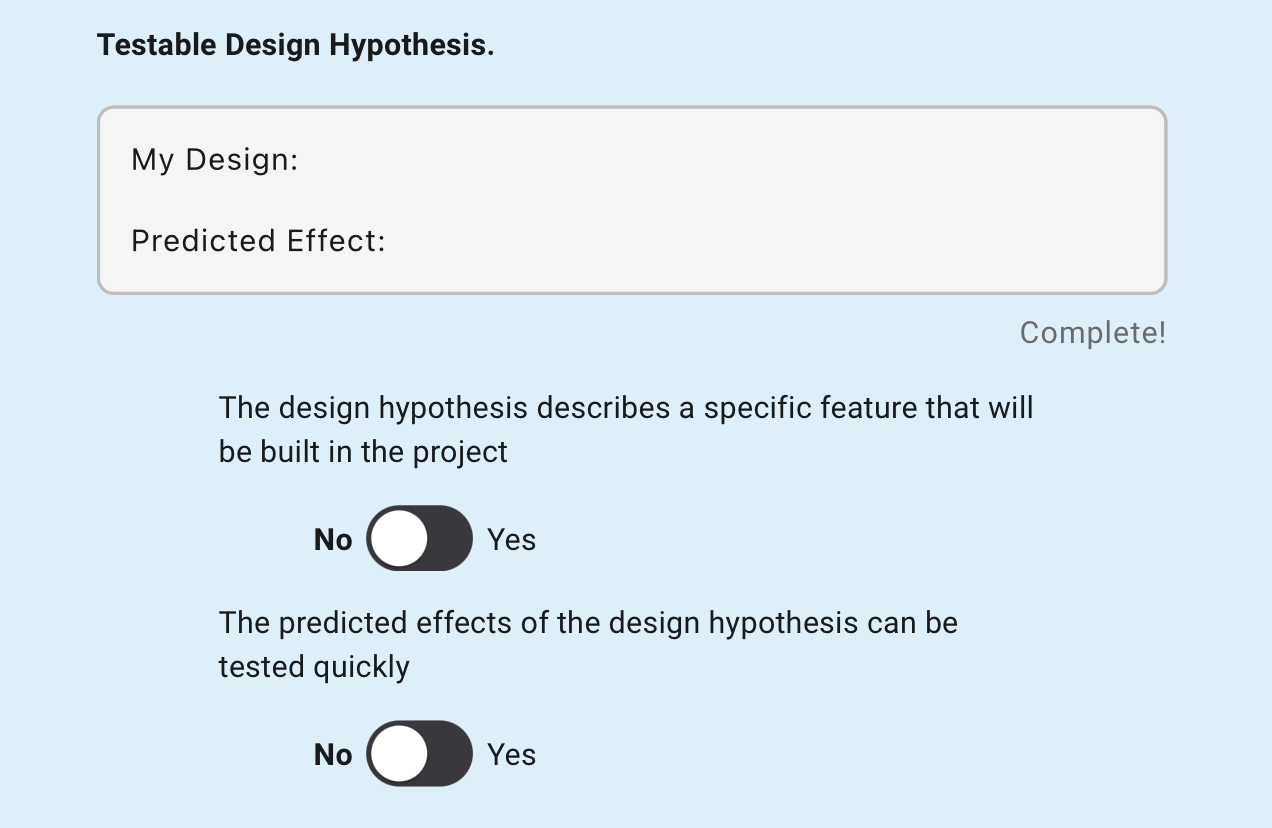}}
\end{figure}

\begin{figure}[t]
\floatconts
  {fig:s9}
  {\caption{The system prompts the student to reflect on their proposal in terms of evaluation plan.}}
  {\includegraphics[width=0.9\textwidth]{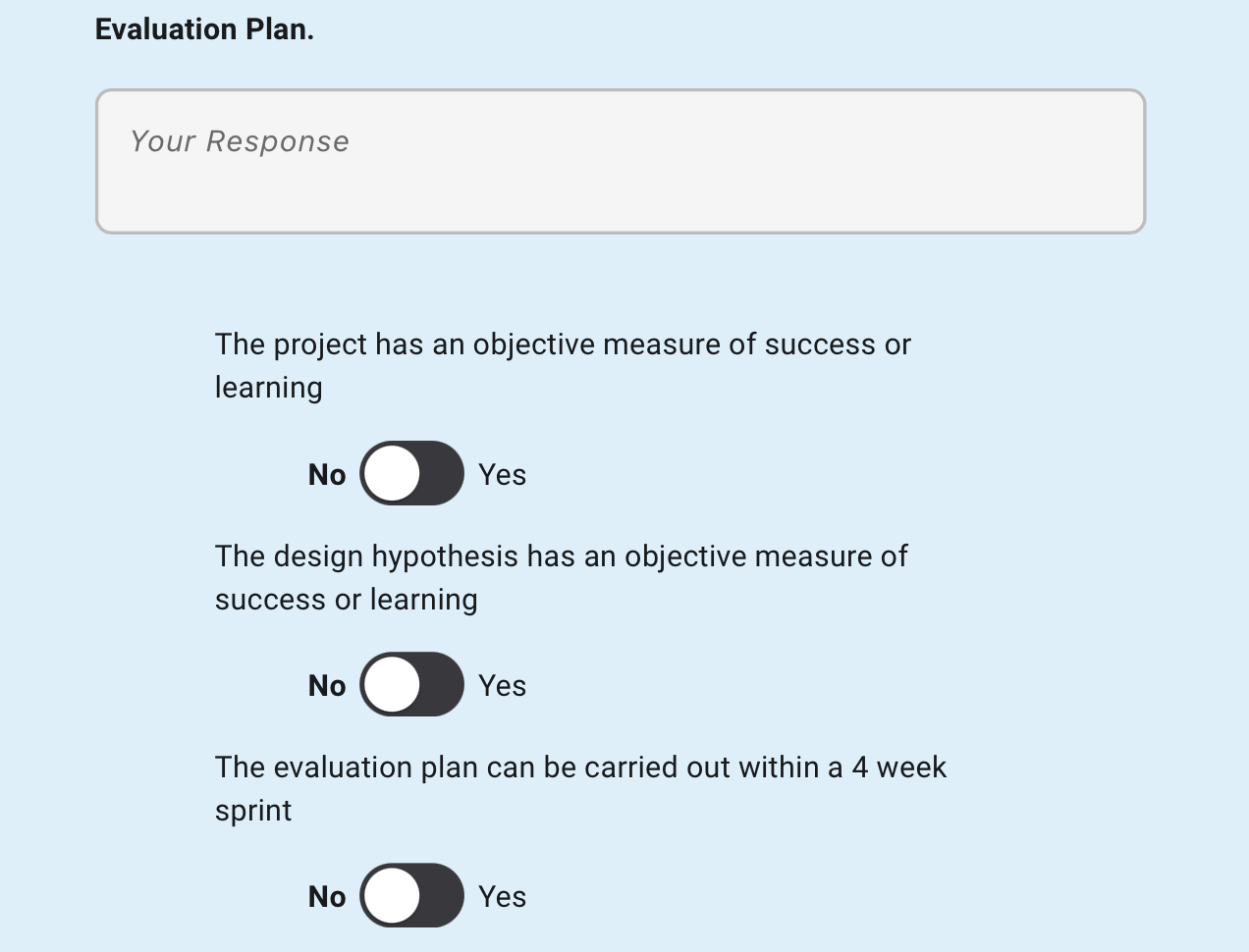}}
\end{figure}

\begin{figure}[t]
\floatconts
  {fig:s10}
  {\caption{The system prompts the student to describe three skills they want to develop while working on their project.}}
  {\includegraphics[width=0.9\textwidth]{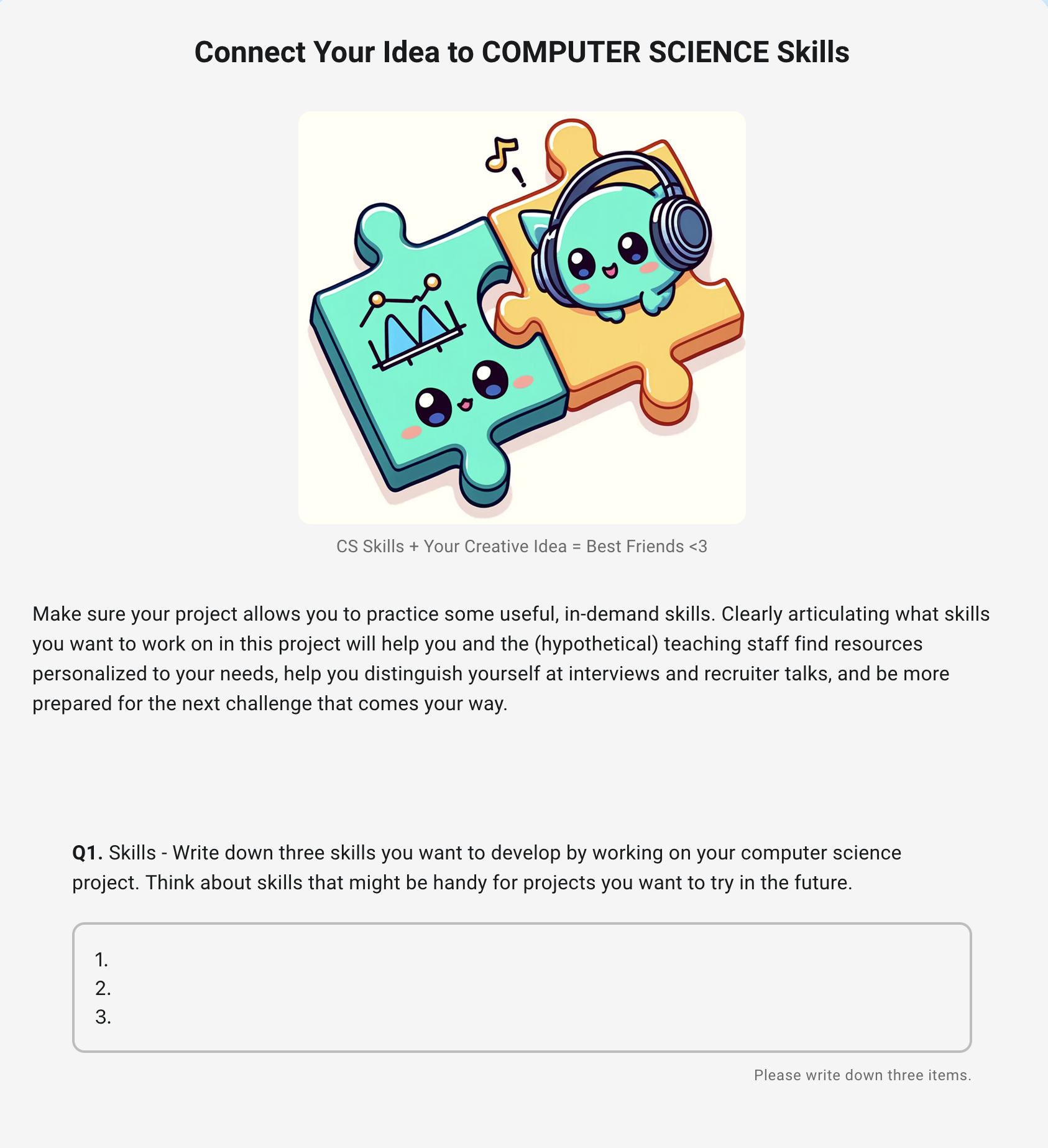}}
\end{figure}

\begin{figure}[t]
\floatconts
  {fig:s11}
  {\caption{The system prompts the student to connect the first skill to a mentor whose perspective would be valuable.}}
  {\includegraphics[width=0.9\textwidth]{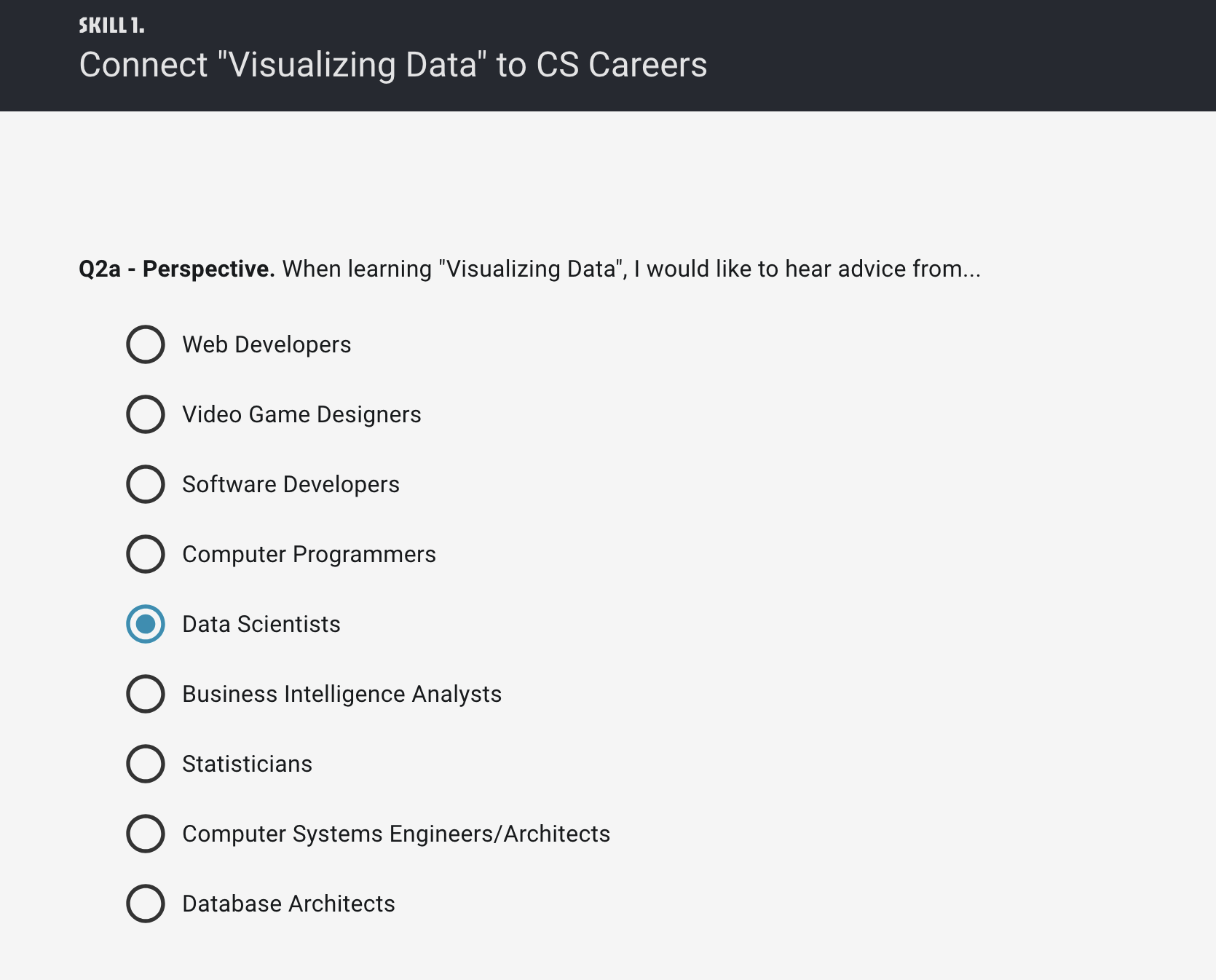}}
\end{figure}

\begin{figure}[t]
\floatconts
  {fig:s12}
  {\caption{The system prompts the student to connect the first skill to one of the top 5 important tasks for their selected mentor's career (tasks sourced from O*Net Online).}}
  {\includegraphics[width=0.9\textwidth]{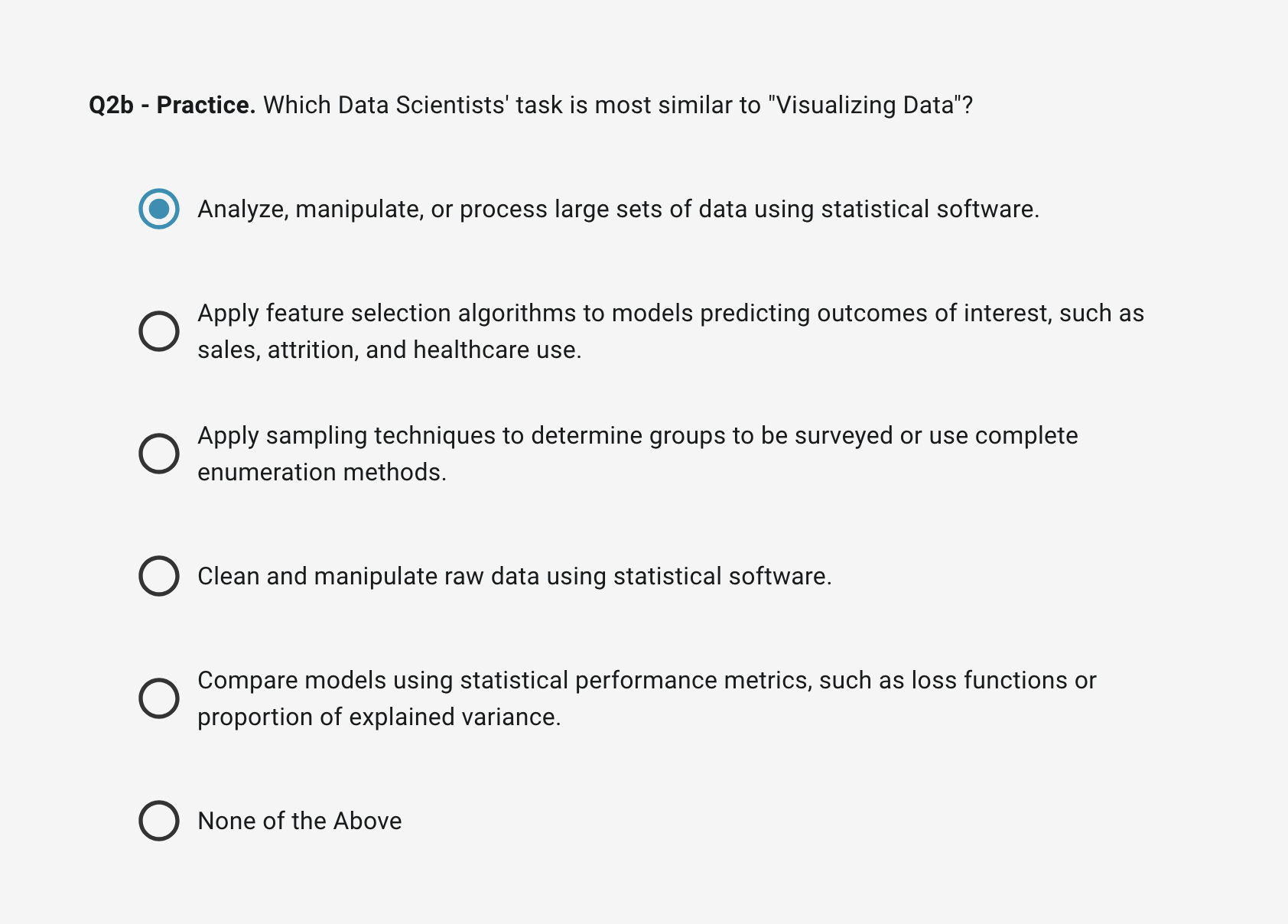}}
\end{figure}

\begin{figure}[t]
\floatconts
  {fig:s13}
  {\caption{The system prompts the student to connect the first skill to tools and technologies associated with their selected mentor's career (technologies sourced from O*Net Online).}}
  {\includegraphics[width=0.9\textwidth]{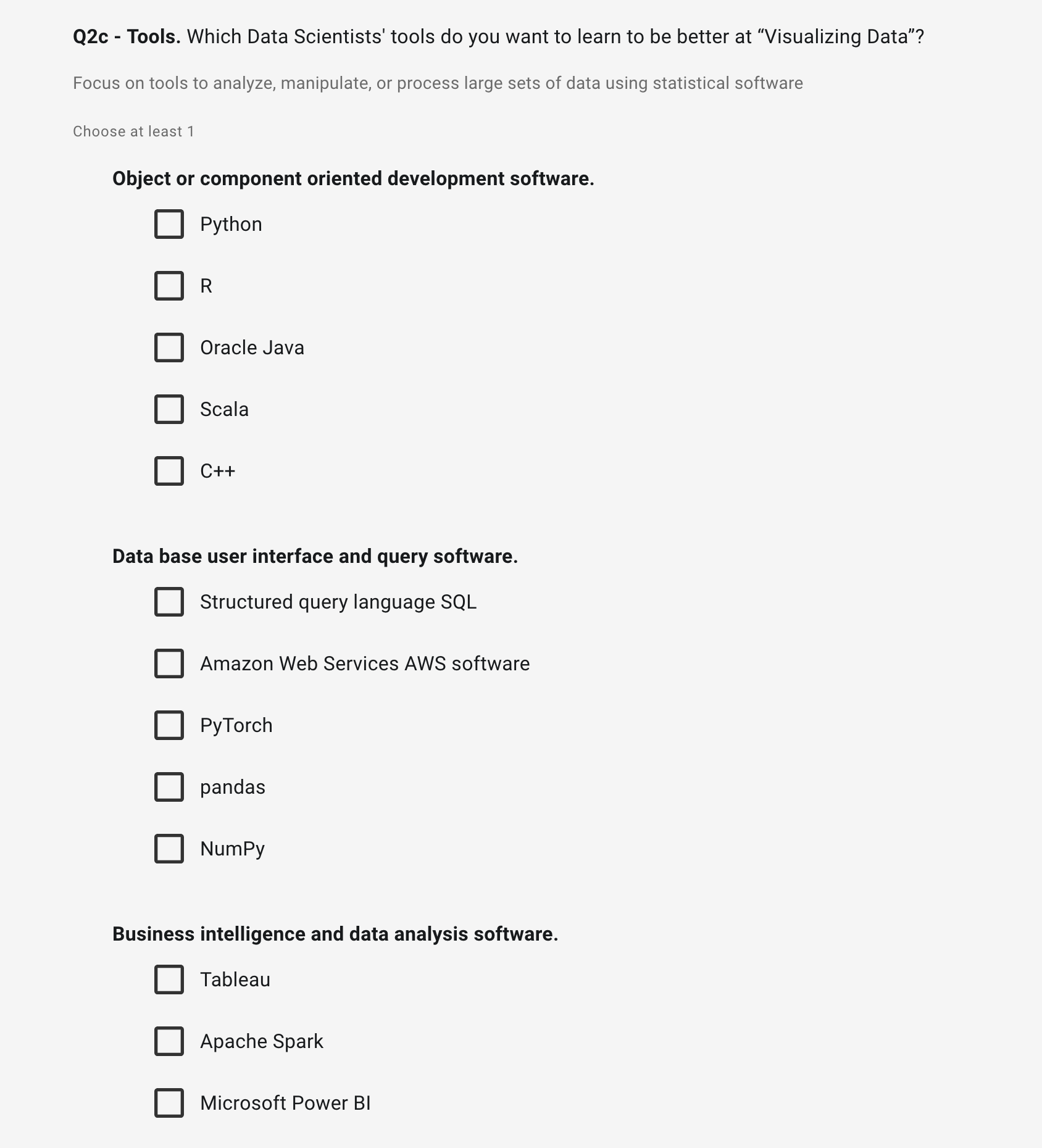}}
\end{figure}

\FloatBarrier

\section{Human Expert Evaluation Rubric}\label{app:expert_evaluation_rubric}

Here, we provide a crowd worker's project proposal and the 23-item quality checklist that human experts and GPT-4o used to grade the project proposals. The rubric had four subtasks: 
\begin{itemize}
    \item Using the same quality checklist students used to self-assess their own work (10 items).
    \item Judging the appropriateness of skill-career pairings (9 items; 3 items per skill).
    \item An overall quality judgment question (“I would recommend a student include this project on their resume”) (1 item).
    \item Judging the quality of skill descriptions written by students (3 items; 1 item per skill). This set of questions was asked in a separate form developed to capture factors that could explain low rater agreement on the appropriateness of skill-career pairings.
\end{itemize}

\begin{figure}[t]
\floatconts
  {fig:e4}
  {\caption{Evaluation task description, shown to the human experts.}}
  {\includegraphics[width=0.85\textwidth]{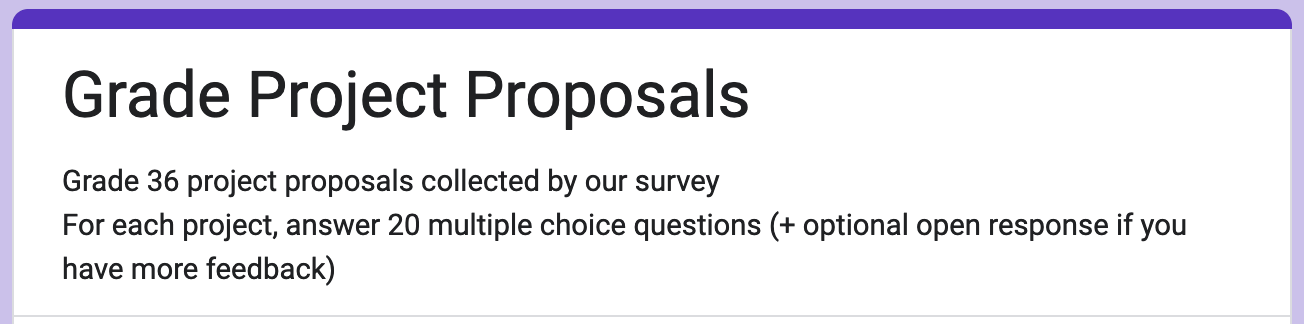}}
\end{figure}

\begin{figure}[t]
\floatconts
  {fig:e5}
  {\caption{Instructions for evaluating project proposals, shown to the human experts.}}
  {\includegraphics[height=0.85\textheight]{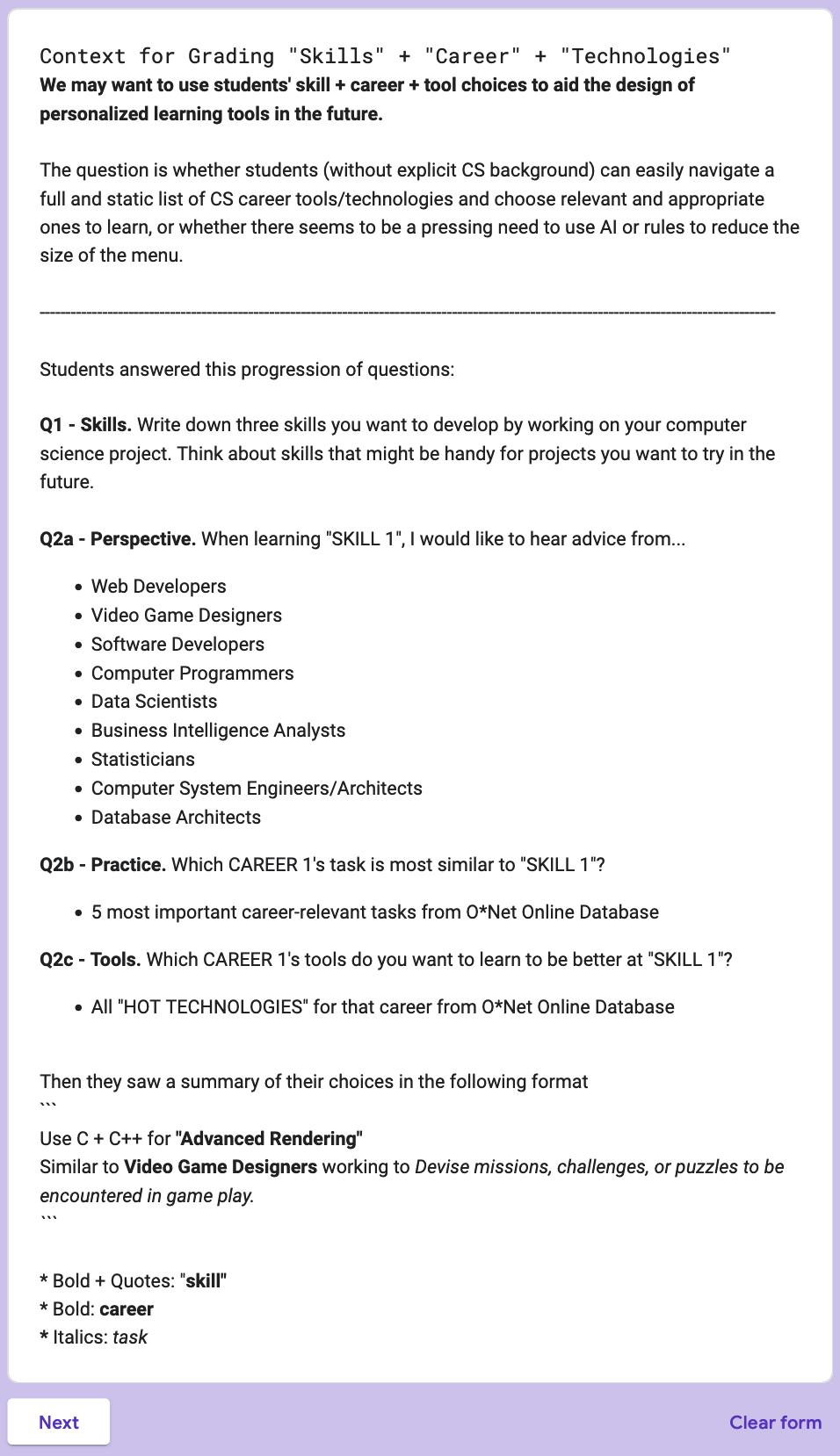}}
\end{figure}

\begin{figure}[t]
\floatconts
  {fig:e6}
  {\caption{Items for evaluating a student's project proposal in terms of background and problem.}}
  {\includegraphics[width=0.85\textwidth]{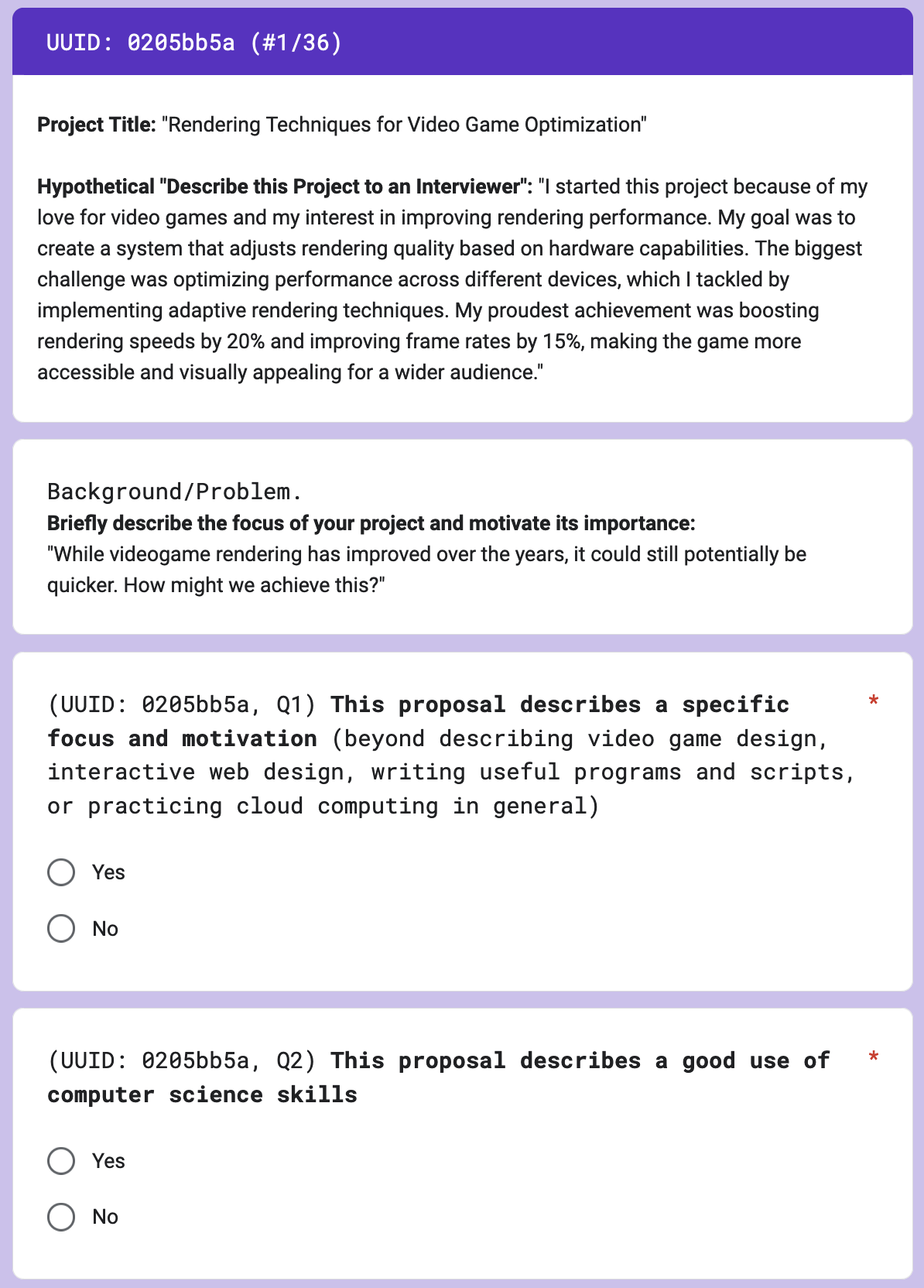}}
\end{figure}

\begin{figure}[t]
\floatconts
  {fig:e7}
  {\caption{Items for evaluating a student's project proposal in terms of objectives.}}
  {\includegraphics[width=0.85\textwidth]{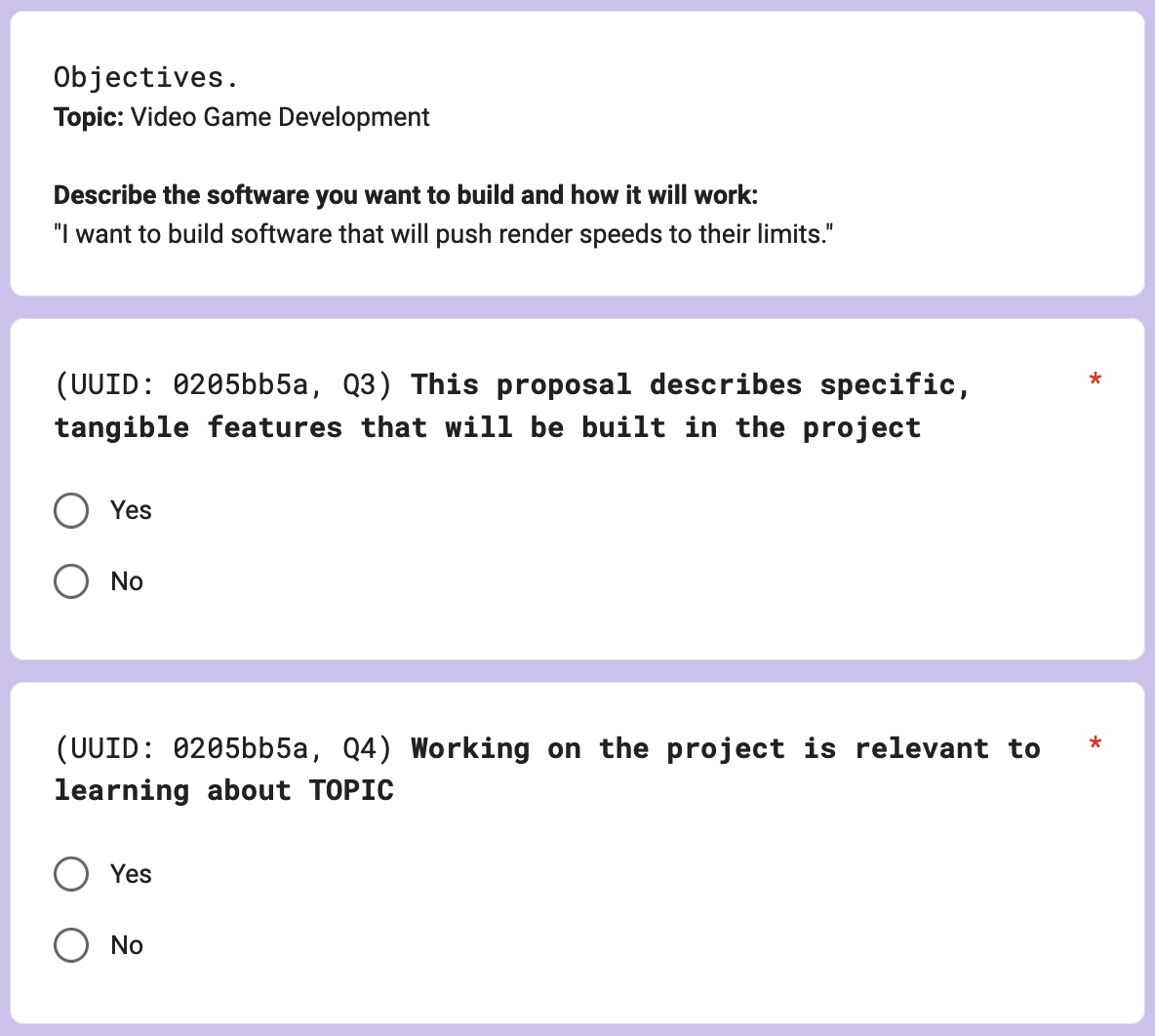}}
\end{figure}

\begin{figure}[t]
\floatconts
  {fig:e8}
  {\caption{Items for evaluating a student's project proposal in terms of related work.}}
  {\includegraphics[width=0.85\textwidth]{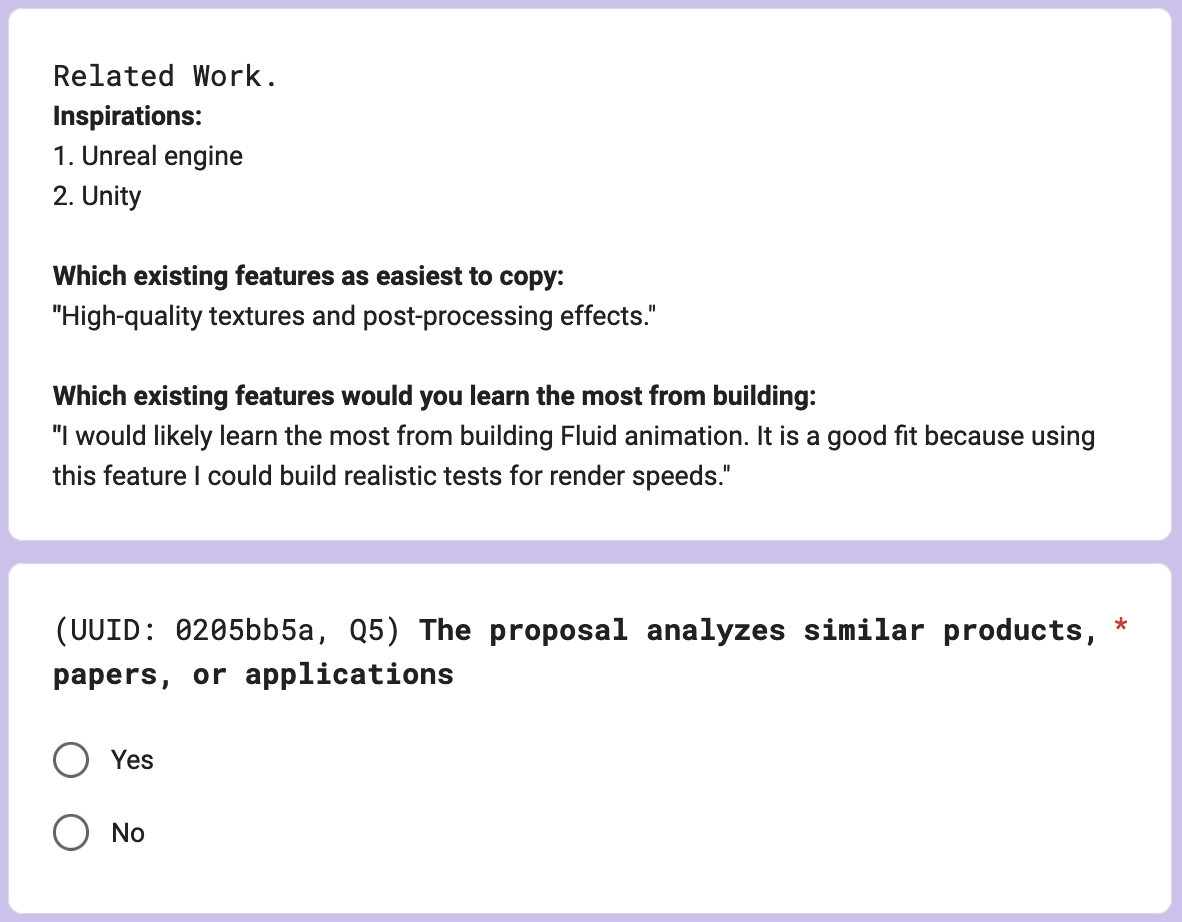}}
\end{figure}

\begin{figure}[t]
\floatconts
  {fig:e9}
  {\caption{Items for evaluating a student's project proposal in terms of design hypothesis.}}
  {\includegraphics[width=0.85\textwidth]{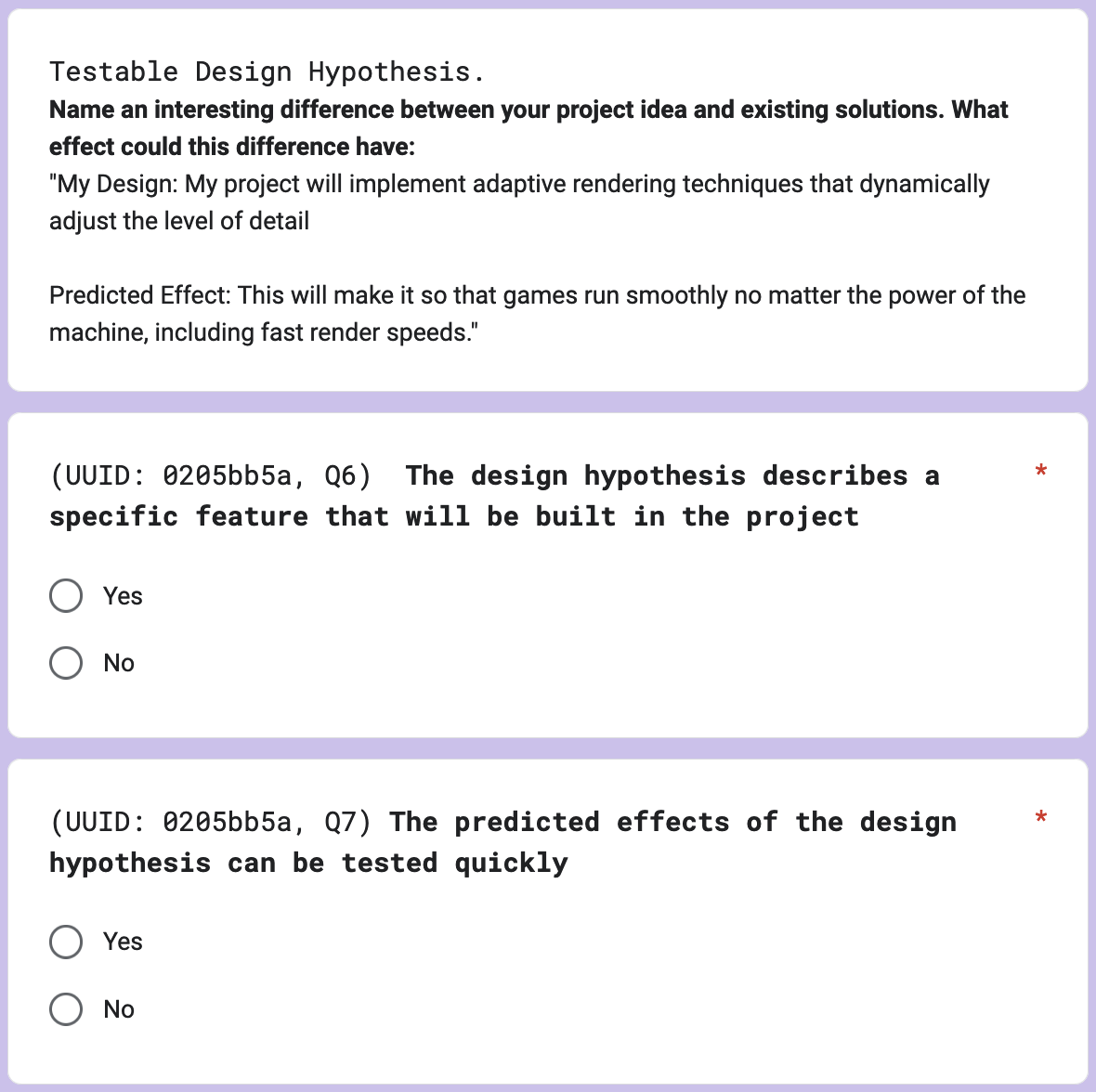}}
\end{figure}

\begin{figure}[t]
\floatconts
  {fig:e10}
  {\caption{Items for evaluating a student's project proposal in terms of evaluation plan.}}
  {\includegraphics[width=0.85\textwidth]{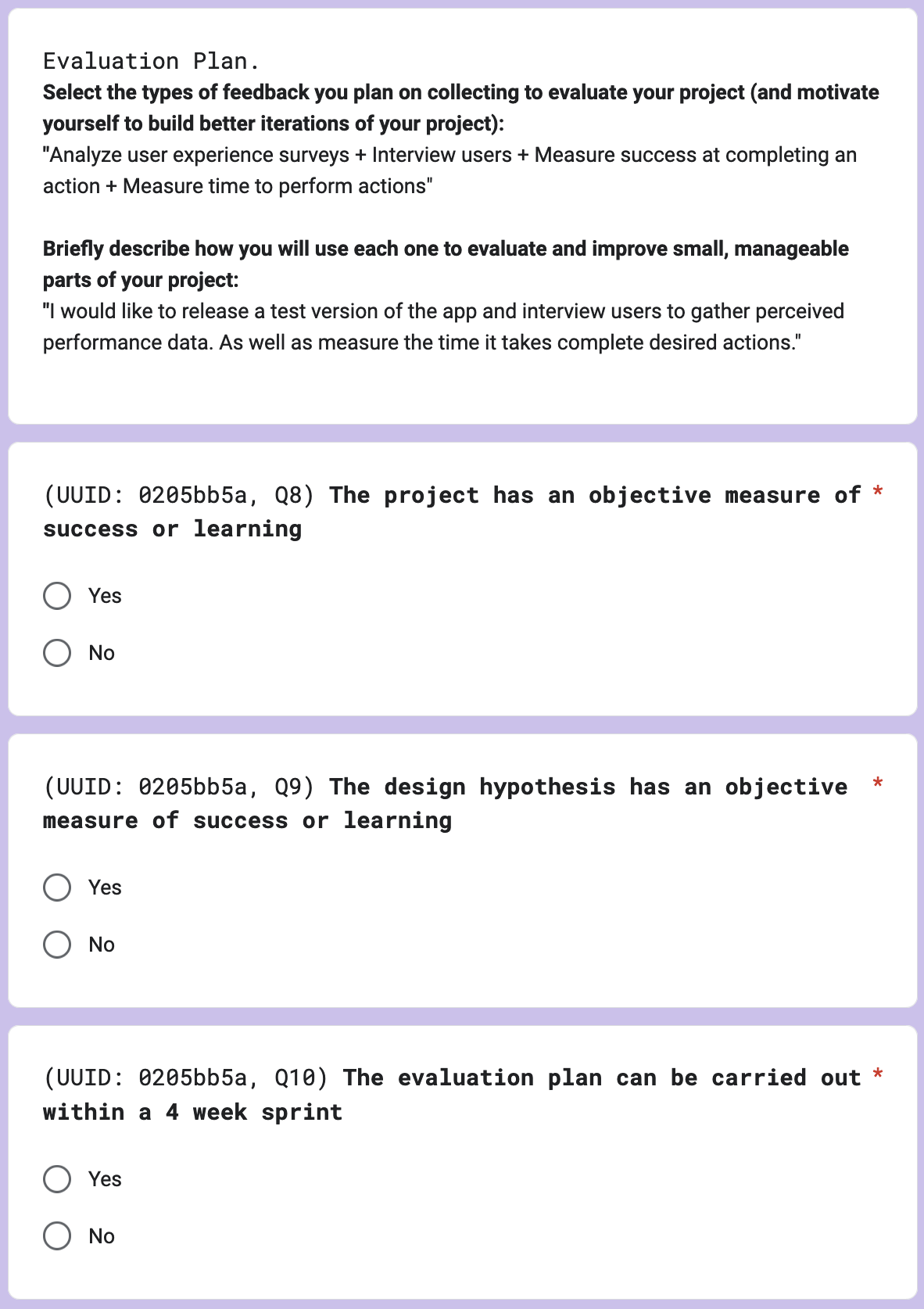}}
\end{figure}

\begin{figure}[t]
\floatconts
  {fig:e11}
  {\caption{Items for evaluating whether student has matched their skill \#1 to appropriate technologies, computer science career, and career task.}}
  {\includegraphics[width=0.85\textwidth]{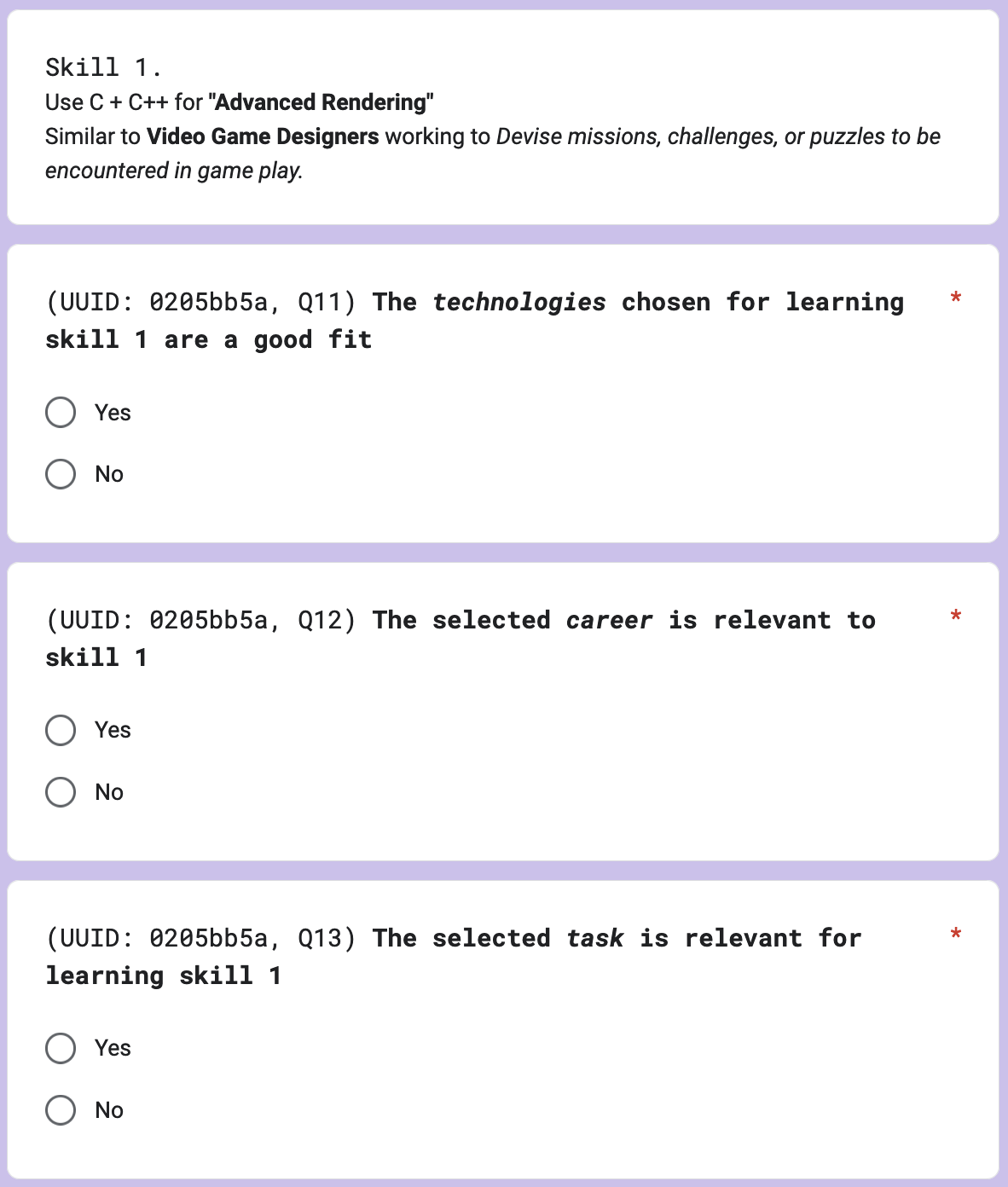}}
\end{figure}

\begin{figure}[t]
\floatconts
  {fig:e12}
  {\caption{Items for evaluating whether student has matched their skill \#2 to appropriate technologies, computer science career, and career task.}}
  {\includegraphics[width=0.85\textwidth]{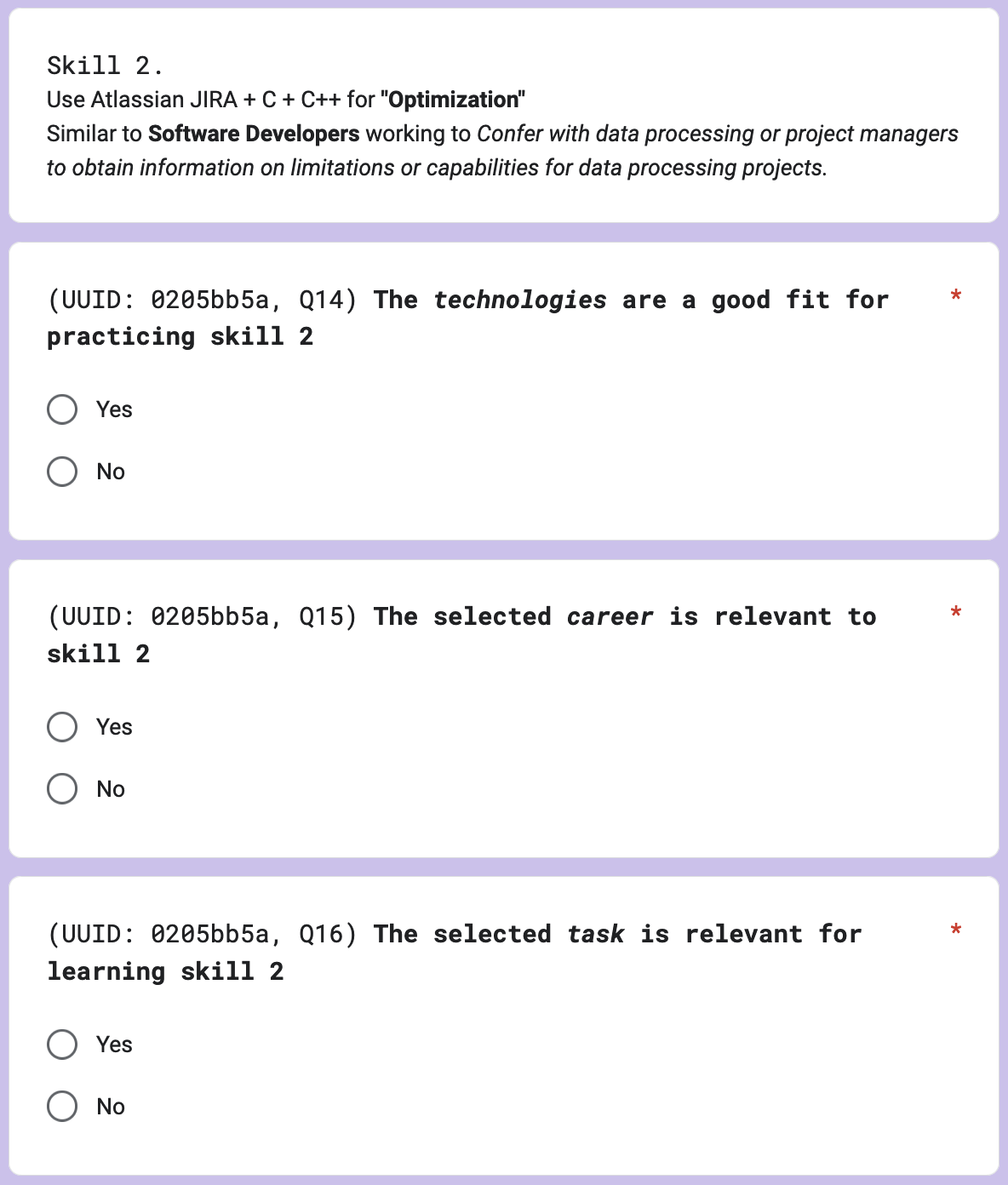}}
\end{figure}

\begin{figure}[t]
\floatconts
  {fig:e13}
  {\caption{Items for evaluating whether student has matched their skill \#3 to appropriate technologies, computer science career, and career task.}}
  {\includegraphics[width=0.85\textwidth]{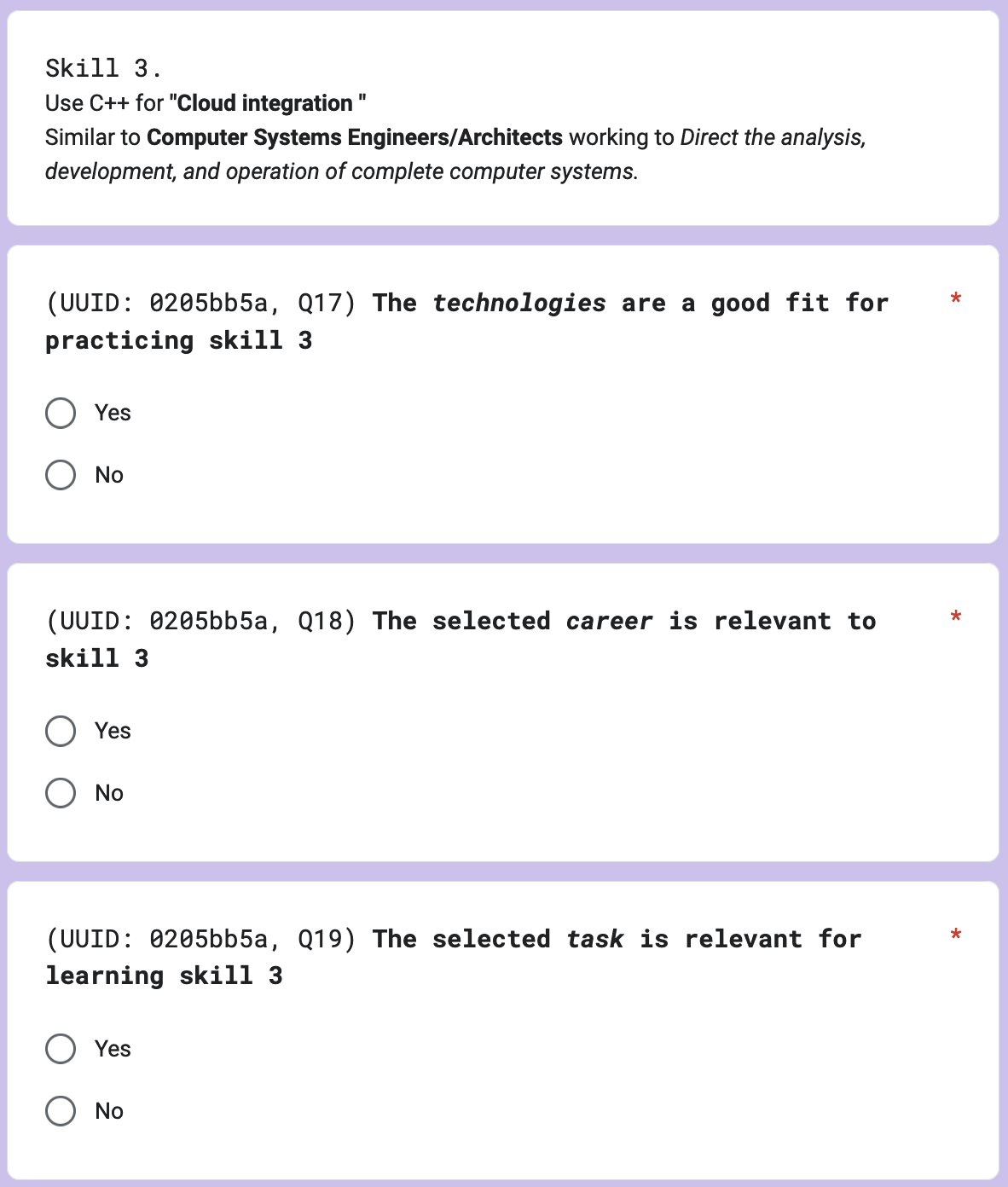}}
\end{figure}

\begin{figure}[t]
\floatconts
  {fig:e14}
  {\caption{Item for evaluating overall quality and open-ended comments on the project proposal.}}
  {\includegraphics[width=0.85\textwidth]{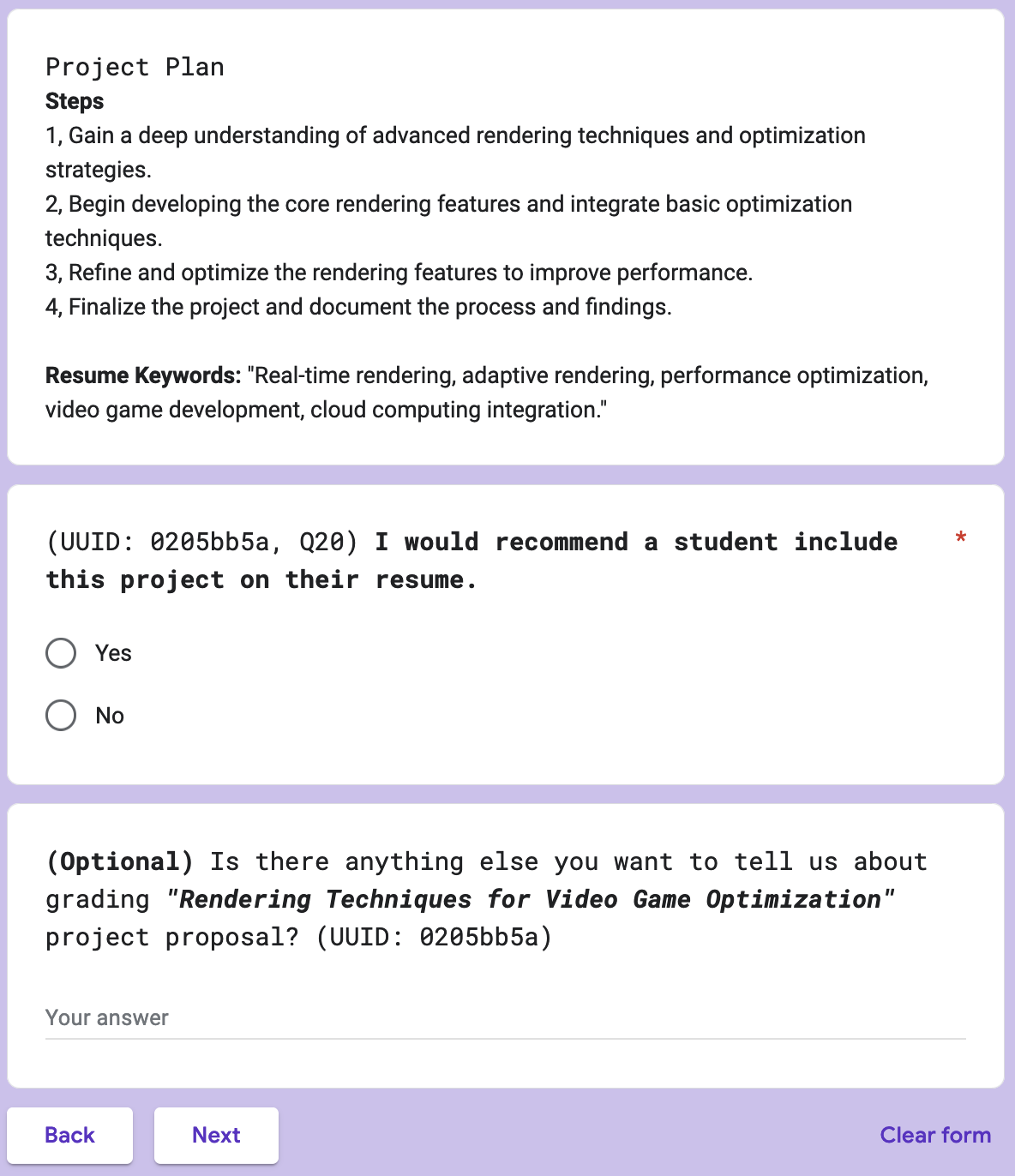}}
\end{figure}

\begin{figure}[t]
\floatconts
  {fig:e1}
  {\caption{Instructions for classifying the skills written by students.}}
  {\includegraphics[width=0.85\textwidth]{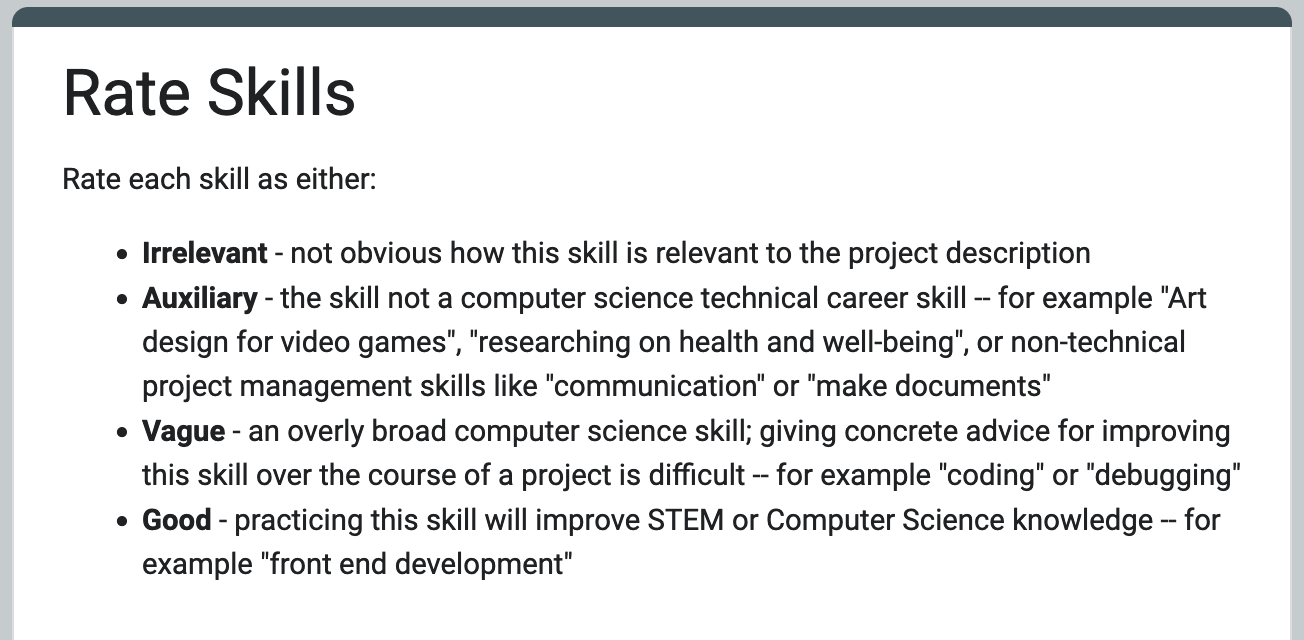}}
\end{figure}

\begin{figure}[t]
\floatconts
  {fig:e2}
  {\caption{Project proposal and skill \#1 classification question.}}
  {\includegraphics[width=0.85\textwidth]{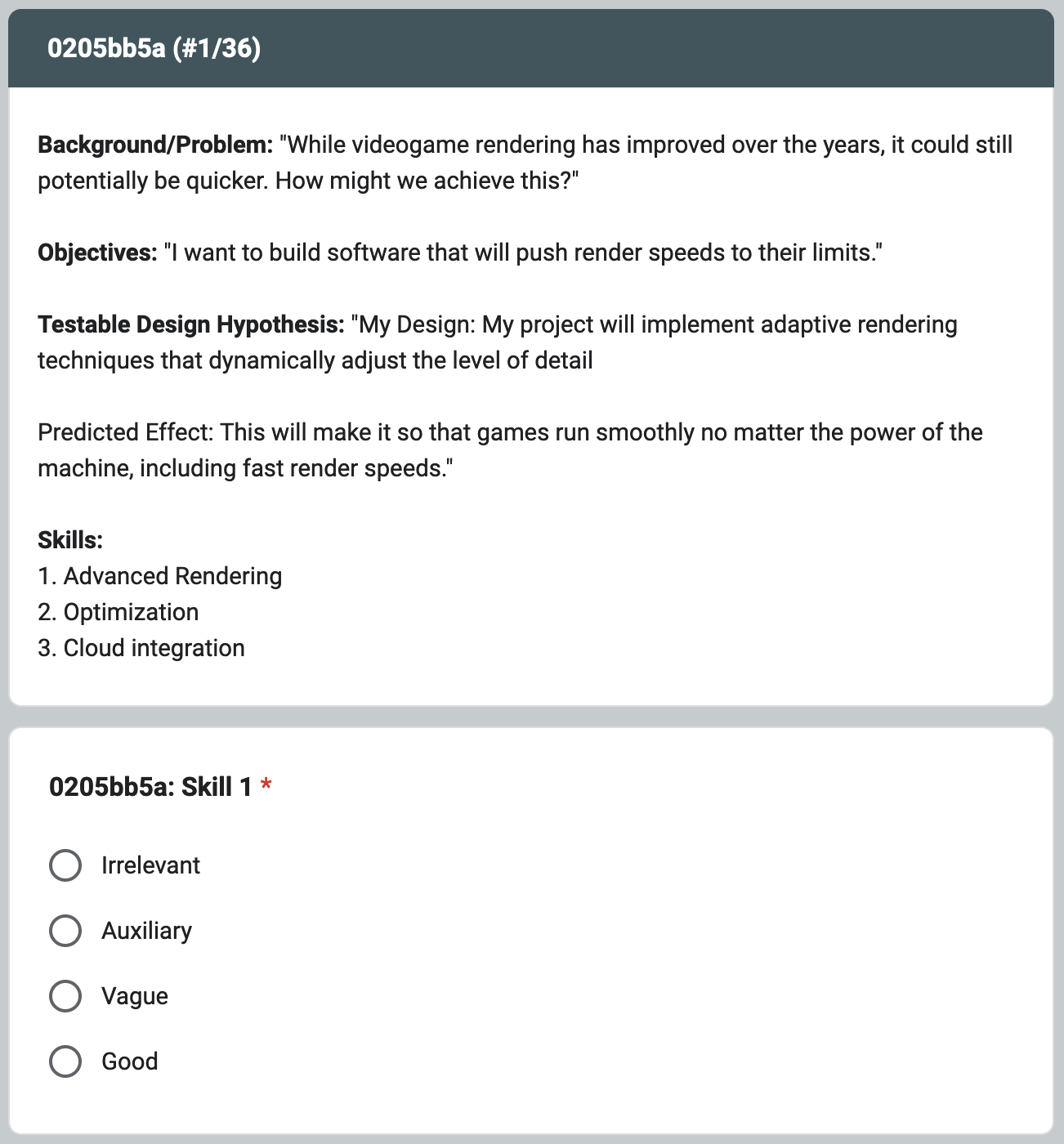}}
\end{figure}

\begin{figure}[t]
\floatconts
  {fig:e3}
  {\caption{The skill \#1 and skill \#2 classification questions.}}
  {\includegraphics[width=0.85\textwidth]{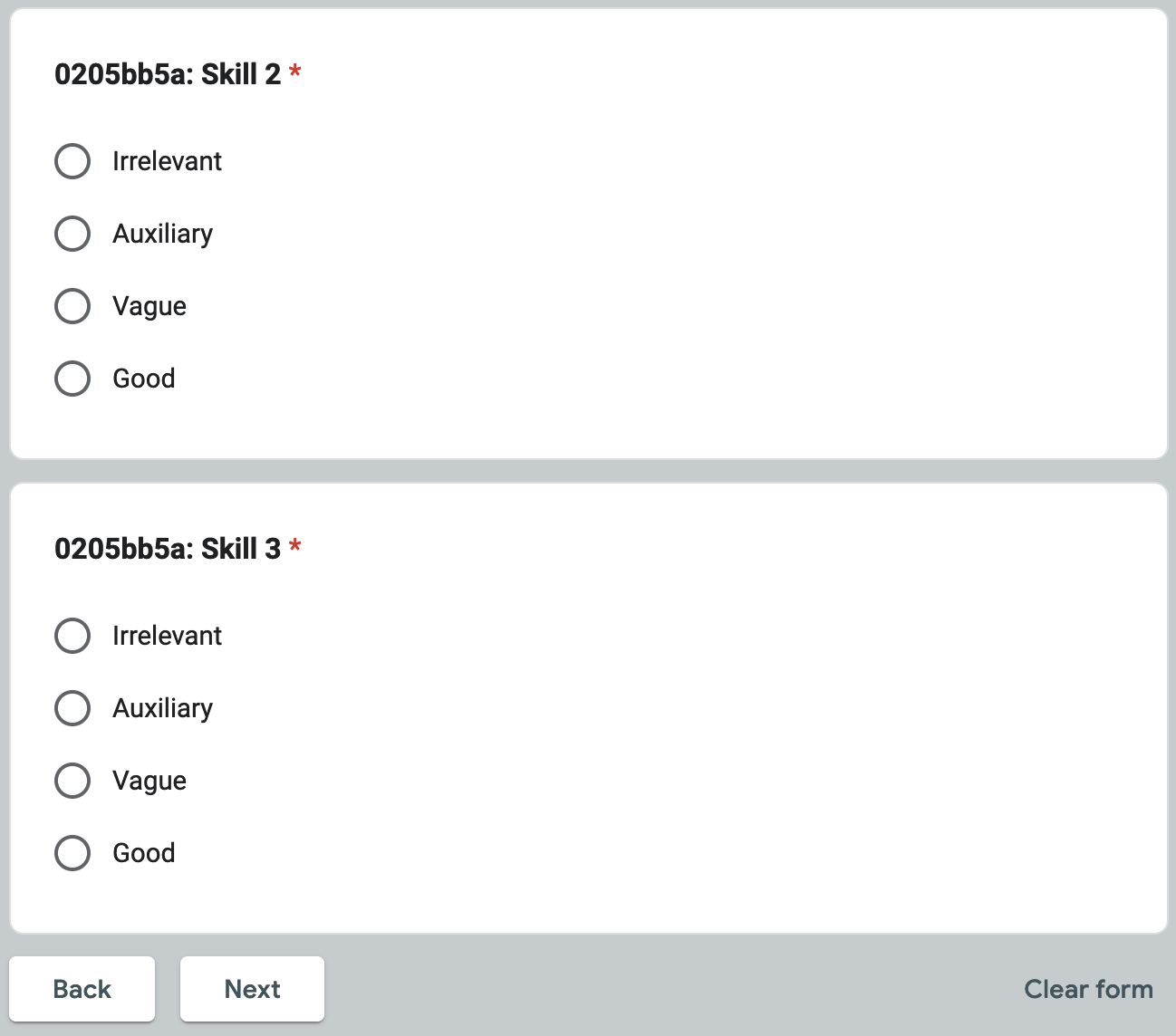}}
\end{figure}

\FloatBarrier

\section{LLM-as-a-Judge Prompt GPT-4o}\label{app:llm-prompts}

To produce ratings from GPT-4o, we call OpenAI API with the following prompt templates, populated with each project proposal's text. The rubric items are the same as the items used in the human expert grader's rubric given in Appendix \ref{app:expert_evaluation_rubric}.

\begin{table}[h!]
\centering
 \begin{tabular}{||p{0.85\textwidth}||} 
 \hline
 System Prompt - Quality Checklist \& Resume Project \\ [0.5ex] 
 \hline\hline
 You are a teaching assistant. Your job is to check project proposals for quality and higher order thinking skills in Bloom's taxonomy (evaluation, synthesis). For each item in the checklist (Q1-Q11), respond with Yes or No. \\ [1ex] 
 \hline
 \end{tabular}
\end{table}

\begin{table}
\centering
 \begin{tabular}{||p{0.95\textwidth}||} 
 \hline
 User Prompt - Quality Checklist \& Resume Project \\ [0.5ex] 
 \hline\hline
Project Title: \`{}\`{}\`{}project\_title\`{}\`{}\`{} \\
Background/Problem. \\
Briefly describe the focus of your project and motivate its importance: \`{}\`{}\`{}q2\_problem\`{}\`{}\`{} \\
* Q1. This proposal describes a specific focus and motivation (beyond describing video game design, interactive web design, writing useful programs and scripts, or practicing cloud computing in general) \\
* Q2. This proposal describes a good use of computer science skills \\ \\

Objectives: \\
Topic: \`{}\`{}\`{}q1\_topic\`{}\`{}\`{} \\
Describe the software you want to build and how it will work: \`{}\`{}\`{}q3\_objective\`{}\`{}\`{} \\
* Q3. This proposal describes specific, tangible features that will be built in the project \\
* Q4. Working on the project is relevant to learning about q1\_topic \\ \\

Related Work. \\
Inspirations: \`{}\`{}\`{}q4\_inspirations\`{}\`{}\`{} \\
Which existing features are easiest to copy: \`{}\`{}\`{}q5\_analysis\_copy\`{}\`{}\`{} \\
Which existing features would you learn the most from building: \`{}\`{}\`{}q6\_analysis\_learn\`{}\`{}\`{} \\
* Q5. The proposal analyzes similar products, papers, or applications \\ \\

Testable Design Hypothesis. \\
Name an interesting difference between your project idea and existing solutions. What effect could this difference have: \`{}\`{}\`{}q7\_your\_design\`{}\`{}\`{} \\
* Q6. The design hypothesis describes a specific feature that will be built in the project \\ 
* Q7. The predicted effects of the design hypothesis can be tested quickly \\ \\

Evaluation Plan. \\
Select the types of feedback you plan on collecting to evaluate your project (and motivate yourself to build better iterations of your project): \`{}\`{}\`{}q8\_feedback\`{}\`{}\`{} \\
Briefly describe how you will use each one to evaluate and improve small, manageable parts of your project: \`{}\`{}\`{}q9\_evaluation\_plan\`{}\`{}\`{} \\
* Q8. The project has an objective measure of success or learning  \\
* Q9. The design hypothesis has an objective measure of success or learning \\
* Q10. The evaluation plan can be carried out within a 4 week sprint \\ \\

Project Plan. \\
Steps: \`{}\`{}\`{}steps\`{}\`{}\`{} \\
* Q11. I would recommend a student include this project on their resume. \\ \\

Desired format: \\
* QX. \{Yes $\mid$ No\} \\ [1ex] 
 \hline
 \end{tabular}
\end{table}

\begin{table}[h!]
\centering
 \begin{tabular}{||p{0.85\textwidth}||}  
 \hline
 System Prompt - Skill Pairing Classification \\ [0.5ex] 
 \hline\hline
 You are a teaching assistant. Your job is to check whether the skill students want to learn is appropriate for the listed career and career-specific task. The skill students want to learn is in double quotes. The mentor and the task are listed on the following line. For each item in the checklist (Q1-Q3), respond with Yes or No. \\ [1ex] 
 \hline
 \end{tabular}
\end{table}

\begin{table}[h!]
\centering
 \begin{tabular}{||p{0.85\textwidth}||}
 \hline
 User Prompt - Skill Pairing Classification \\ [0.5ex] 
 \hline\hline
Background/Problem: \`{}\`{}\`{}q2\_problem\`{}\`{}\`{} \\

Objectives: \`{}\`{}\`{}q3\_objective\`{}\`{}\`{} \\

Testable Design Hypothesis: \`{}\`{}\`{}q7\_your\_design\`{}\`{}\`{} \\

Skill: \`{}\`{}\`{}skill\`{}\`{}\`{} \\
* Q1. The technologies chosen for learning the skill are a good fit. \\
* Q2. The selected career is relevant to the skill. \\
* Q3. The selected career-specific task is relevant for learning the skill. \\

Desired format:
* QX. \{Yes $\mid$ No\} \\ [1ex] 
 \hline
 \end{tabular}
\end{table}

\begin{table}[h!]
\centering
 \begin{tabular}{||p{0.85\textwidth}||}
 \hline
 System Prompt - Skill Classification \\ [0.5ex] 
 \hline\hline
 You are a teaching assistant. Rate each skill as either: \\
* Irrelevant - not obvious how this skill is relevant to the project description \\
* Auxiliary - the skill not a computer science technical career skill -- for example ``Art design for video games", ``researching on health and well-being", or non-technical project management skills like ``communication" or ``make documents" \\
* Vague - an overly broad computer science skill; giving concrete advice for improving this skill over the course of a project is difficult -- for example ``coding" or ``debugging" \\
* Good - practicing this skill will improve STEM or Computer Science knowledge -- for example ``front end development" \\
Respond with Irrelevant, Auxiliary, Vague, or Good. \\ [1ex] 
 \hline
 \end{tabular}
\end{table}

\begin{table}[h!]
\centering
 \begin{tabular}{||p{0.85\textwidth}||} 
 \hline
 User Prompt - Skill Classification \\ [0.5ex] 
 \hline\hline
 Background/Problem: \`{}\`{}\`{}q2\_problem\`{}\`{}\`{} \\

Objectives: \`{}\`{}\`{}q3\_objective\`{}\`{}\`{} \\

Testable Design Hypothesis: \`{}\`{}\`{}q7\_your\_design\`{}\`{}\`{} \\

Skill: \`{}\`{}\`{}skill\`{}\`{}\`{} \\
Desired format: \\
\{Irrelevant $\mid$ Auxiliary $\mid$ Vague $\mid$ Good\} \\ [1ex] 
 \hline
 \end{tabular}
\end{table}

\FloatBarrier

\section{Recruitment Call Posted on Prolific}
\label{app:prolific_recruitment}

\begin{figure}[t]
\floatconts
  {fig:prolific}
  {\caption{Recruitment call displayed on Prolific crowd worker platform.}}
  {\includegraphics[width=0.85\textwidth]{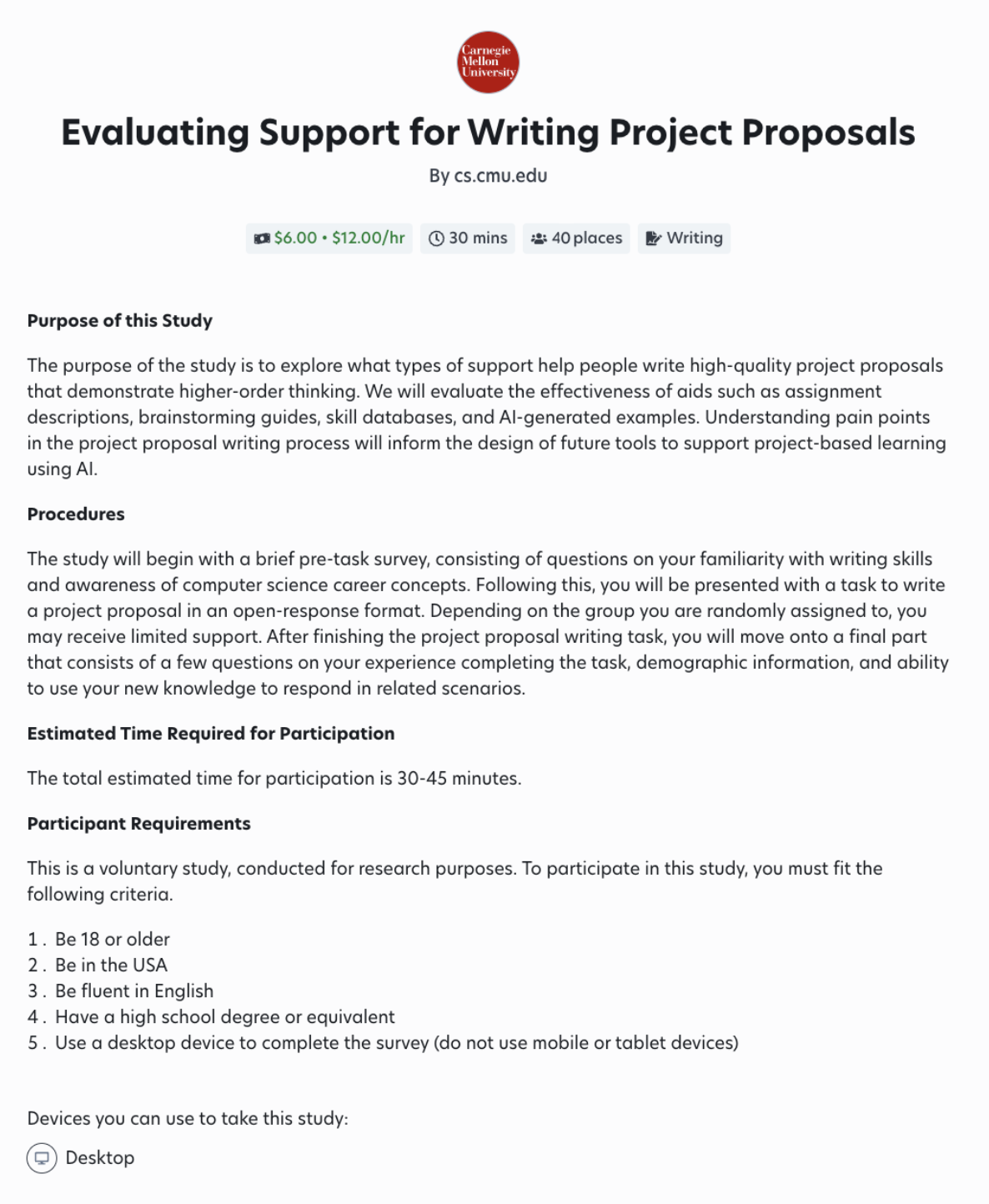}}
\end{figure}